\documentclass[twocolumn,prb,aps,showpacs,amsmath,amssymb,floatfix]{revtex4-1}

\usepackage{amsmath,amssymb,bm,epsfig}
\usepackage{dcolumn}
\usepackage{bm}
\usepackage{feynmf}

\begin{document}

\title{Theory of Phonon-Assisted Adsorption in Graphene: Many-Body Infrared Dynamics}

\author{Sanghita Sengupta}

\affiliation{Institut Quantique and D$\acute{e}$partement de Physique, Universit$\acute{e}$ de Sherbrooke, Sherbrooke, Qu$\acute{e}$bec, Canada J1K 2R1}

\date{\today}
\begin{abstract}
    We devise a theory of adsorption of low-energy atoms on suspended graphene membranes maintained at low temperature based on a model of atom-acoustic phonon interactions. Our primary technique includes a non-perturbative method which treats the dynamics of the multiple phonons in an exact manner within the purview of the Independent Boson Model (IBM). We present a study on the effects of the phonons assisting the renormalization as well as decay of the incident atom propagator and discuss results for the many-body adsorption rates for atomic hydrogen on graphene micromembranes. Additionally, we report similarities of this model with other branches of quantum field theories that include long-range interactions like quantum electrodynamics (QED) and perturbative gravity. 
\end{abstract}
\maketitle
\section{Introduction}
\label{sec:intro}
How does an atom adsorb to a surface? From the viewpoint of quantum field theory, can we understand this phenomenon by including the effects of the surface phonons? Over the years, theoretical predictions as well as experimental endeavors have elucidated a significant role played by the surface phonons in mediating the process of adsorption\cite{Brenig1979,Böheim1982,Brenig1980,Sedlmeir1980,schlichting,jackson}. While most of the previous work has been conducted on conventional three-dimensional materials, the discovery of graphene has led to a recent effort to understand this phenomenon with special focus on phonon dispersion, tunability of the atom-phonon interaction and possible applications\cite{Bonfanti_2018,bruno,DPC2013,Sengupta2016,Sengupta2017}. This brings us to the topic of our current work. 

In this paper, we will devise a theory of adsorption based on a simple model of atom-phonon interaction. Using the tools of quantum field theory, we predict adsorption rates for these suspended graphene membranes that are maintained at low temperature. Let us also mention that our naive model shows an interesting similarity with other branches of quantum field theory. Theories with long-range interactions like quantum electrodynamics and perturbative gravity are seen to exhibit severe infrared (IR) divergences in their scattering rates due to the emission of infinitely many soft quanta (soft meaning vibrations with energy $\epsilon\rightarrow 0$)\cite{Bloch:1937pw,landaulifshitz4,weinberg,weinberg_1995}. Quite remarkably, our non-relativistic model is also plagued with severe IR divergences in the adsorption rate that appear as a result of emission of infinitely many soft phonons originating from the long-range tail of van der Waals (vdW) interactions\cite{Sengupta2016,Sengupta2017,sengupta2018}. Since physically measurable entities can never be infinity, this IR-divergent adsorption rate poses a serious concern for the application and validity of the theory. However, thanks to the Kinoshita-Lee-Nauenberg theorem\cite{kinoshita,lee}, we realize that these infinities are infact unreal and proper application of resummation procedures can lead us to meaningful results. 

Quite naturally there is an ongoing attempt to devise non-perturbative methods to tackle these IR divergences\cite{Sengupta2016,Sengupta2017,clougherty2017,clougherty2017T}. As a matter of fact, these resummation methods implemented in these condensed matter systems are similar in essence to the corresponding QED counterparts. To illustrate this point, let us briefly mention the main idea behind each of them: (i) Bloch-Nordsieck scheme - the method tackles the IR divergence by allowing for the inclusion of emission of infinitely many soft quanta and summing over them\cite{Bloch:1937pw,qed,qed1,jakovac,jakovac2,jakovac3,Sengupta2016,Sengupta2017,qed,qed1}, (ii) Faddeev-Kulish mechanism - this method proposes a dressing of asymptotic states by a cloud of soft quanta using a coherent state formalism\cite{clougherty2017,clougherty2017,Kulish1970,kibble2,kibble3,kibble4} and (iii) imposing a IR cut-off\cite{LJreply,LJ2011}. While all three methods give comparable and reasonable answers for zero-temperature adsorption rates in graphene micromembranes\cite{Sengupta2017}, the low-temperature result seems contentious with the core of the debate surrounding the effect of the IR cut-off, i.e the effect of low-energy phonons.

In one numerical study performed for low temperature membranes (T= 10 K) \cite{LJ2011,LJreply} authors considered adsorption as a fast process mediated by a single phonon and derived finite, enhanced adsorption rates. To keep their computations tractable, they imposed an IR cut-off. Ref.~[\onlinecite{clougherty2017T}] used a coherent-state phonon basis formalism (dressing the asymptotic states by a cloud of soft phonons), claimed to cure and remedy the IR problem at finite temperature and predicted adsorption rates that tend to zero. However, this method sparks some serious questions. (i) In the derivation of the final adsorption rate $\Gamma$ (obtained by summing over partial rates $\Gamma_{n}$), Ref.~[\onlinecite{clougherty2017T}] has considered emission of soft phonons with energy $\epsilon\sim\omega_{c}$
where $\omega_{c} = 0.183$ meV. Since the definition of IR problem corresponds to $\epsilon \rightarrow 0$, it is not clear to us how the IR problem is remedied in this case. (ii) The partial rates corresponding to \textit{n} phonon processes were derived under the assumption, $\omega_{c}\gg g_{b}^{2}$ with $g_{b}^{2}$ as the atom-phonon coupling strength in the membrane. By previous line of thought this implies $\epsilon\gg g_{b}^{2}$. We are not certain if this is the right energy scale for the problem at hand (presumably, the IR scale should be the lowest energy scale in the formalism).

In light of these recent developments and questions, we are motivated to reconsider this problem in terms of a partial resummation technique that has been previously used for graphene membranes maintained at zero and high temperature (by high, we mean a temperature scale which is comparable to the Debye frequency of the phonons). This technique is based on the Bloch-Nordsieck scheme of resummations and uses the Independent Boson model (IBM) to account for the dynamics of the phonons. While in the zero temperature case, the technique cures the IR problem at $\epsilon\rightarrow0$, for micromembranes it predicts an adsorption rate that is finite, non-zero and size-independent and is within good approximation (1$\%$) of the zero temperature Golden rule result\cite{Sengupta2017}. However, in the high temperature regime, the technique retains some residues of the log singularity because of the temperature effects of the Bose distribution. As a consequence, the method predicts adsorption rates that increase with increasing temperature and membrane sizes. The rate is also enhanced with respect to the finite temperature Golden rule result\cite{Sengupta2016}. 

For the low-temperature formalism, we will not claim to cure the IR problem in this model, in fact we will take a modest approach pertinent to all condensed matter systems. We remind ourselves that the natural IR cut-off is related to the size of the system $\epsilon\sim v_{s}/L$ ($v_{s}$ is the velocity of sound in the material and $L$ is the size of the membrane, we set $\hbar =1$ all through the paper). We thus subject our partial resummation technique to IR cut-offs as low as $\epsilon = 10^{-4}$ meV (the current lowest within all recent literature for low temperature formalism) and systematically increase it to 0.8 meV, naturally, these cut-offs correspond to membrane sizes 10 $\mu$m to 5 nm, respectively. Our primary aim will be to understand how the soft phonons i.e, the IR cut-off, affect the adsorption process with clear focus on the renormalization of the energy of the atom leading to the formation of an acoustic-like polaron and finally the decay of the atom. We will investigate in detail the true dynamics of the phonon bath with respect to the effects of time-evolution, temperature and atom-phonon coupling. To the best of our knowledge, the properties of the true dynamics of the phonon bath related to this model, have not been explored before. As we will see, a key result of this investigation will lead us to a characteristic time scale for the phonon dynamics. This characteristic time scale will indeed serve as a crucial parameter for the prediction of the adsorption rate.

We must also mention that in addition to the resummation technique within the IBM there exists another non-perturbative method, namely the $\textit{Non-Crossing Approximation} (NCA)$ which deals with summing infinite numbers of rainbow diagrams (Feynman diagrams where phonon lines donot cross). This resummation technique was previously found to be successful in treating the IR problem of the model for the zero temperature case and predicted similar result for the adsorption rate as the method of IBM \cite{Sengupta2017}. However for finite temperatures, NCA fails to give tractable convergent results with increasing membrane sizes ($\epsilon\rightarrow0$). Hence as a matter of fact, for the low temperature formalism we will not use the method of NCA but explore the technique within the IBM. The general structure of the paper is as follows: in Sec.~\ref{sec:prelim}, we introduce the model Hamiltonian, the physical features of the model and the methodology to calculate the resummed many-body adsorption rates. Sec.~\ref{sec:IBM} discusses the low-temperature dressed propagator within the IBM, while consequences of atom-phonon interaction on the renormalization and decay of the atom propagator are discussed in Sec.~\ref{sec:phononcorr}. Finally, in Sec.~\ref{sec:MBA}, we derive the many-body adsorption rates based on the partial resummation technique given in Sec.~\ref{sec:prelim} and discuss the results within the context of the phonon-effects explained in Sec.~\ref{sec:phononcorr}. After a summary of our results in Sec.~\ref{sec:conclusion}, we conclude with a few pertaining questions related to this model and IR divergences in other field theories .

\begin{figure}
    \centering
    \includegraphics[width=\columnwidth]{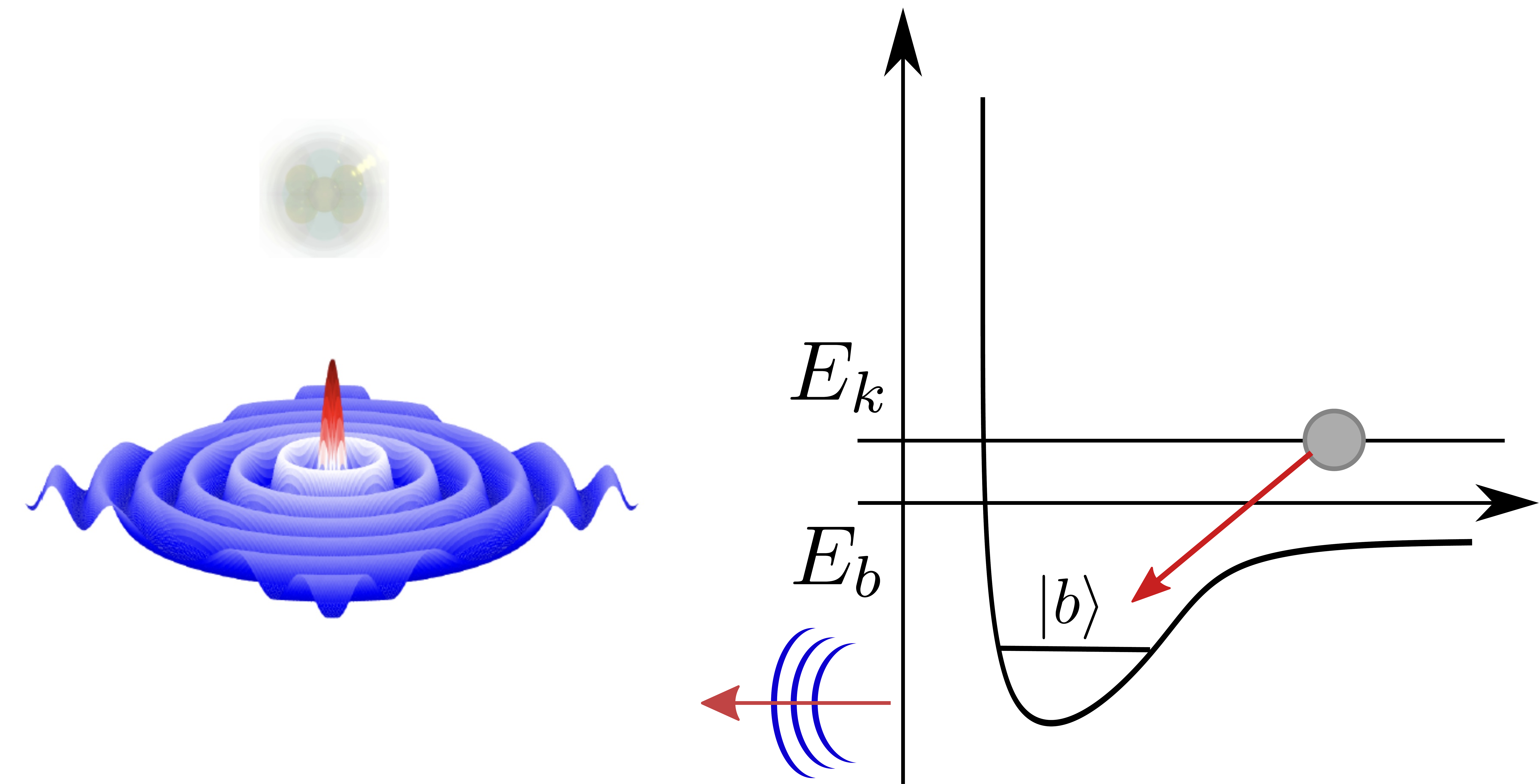}
    \caption{Left: Model of $atom-membrane$ interaction \cite{Sengupta2016,Sengupta2017,clougherty2017}. Weak forces of van der Waals interaction hold together the atom and the membrane that are now separated from each other by a minute distance ($z$). The $\textit{quantized}$ vibrations of the membrane that appear as ripples in the membrane start to interact with the incoming adatom and mediate the process of adsorption \cite{clougherty2017,sengupta2018}. Right: A quantum mechanical description of the adsorption process. Transition of the atom from the continuum $E_{k}$ to the bound state $E_{b}$ (supported in the vdW potential V$_{0}(z)$ of the physisorption well) with the emission of phonons. The Hamiltonian corresponding to the process is explained in Eqs.~\ref{H0}, \ref{Hki} and \ref{Hbi} within Sec.~\ref{sec:prelim}.}
    \label{fig:model}
\end{figure}

\section{Preliminaries and Methodology}
\label{sec:prelim}
In this section, we provide an overview of the physical features of the model and lay out the procedure to calculate the adsorption rate for the atom. 

We consider a continuum model for low-energy physisorption on a suspended graphene membrane. When a low-energy incident atom impinges normally on the graphene membrane, it excites the out-of-plane transverse acoustic mode (ZA) \cite{falkovskii1,LJ2011,Karlický2014} of the membrane thereby mediating interactions that we refer as the \textit{atom-phonon} coupling. Let us begin with a description of the Hamiltonian of our model, $H= H_{0} + H_{i}$, where $H_{0}$ and $H_{i}$ represent the unperturbed and interaction Hamiltonians. For the unperturbed part $H_{0}$, we can write:
\begin{equation}\label{H0}
    H_{0}=  E_{k}c_{k}^{\dagger}c_{k}-E_{b}b^{\dagger}b + \sum_{q}\omega_{q}a_{q}^{\dagger}a_{q}
\end{equation}
where, $E_{k}$ is the energy of the incident atom with the corresponding operators: c$_{k}$ $(c_{k}^{\dagger})$ that annihilates (creates) a particle in the continuum channel;  $b$ $(b^{\dagger})$ annihilates (creates) a particle in the bound state $|b\rangle$ with energy -$E_b$ in the static van der Waals potential $V_{0}(z)$. For graphene membranes, the physisorption well is $E_{b} = 40$ meV \cite{LJ2011,Sengupta2016,Sengupta2017}.
The isotropic surface is modelled by a phonon bath with energy $\omega_q$ \cite{Sengupta2016,Sengupta2017,clougherty2017,Clougherty2012} and operators $a_q$ $(a_q^{\dagger})$ that annihilates(creates) a ZA phonon in the bath \cite{Sengupta2016,Sengupta2017,clougherty2017,Clougherty2012}. For the graphene membrane under an out-of-plane tension $\gamma$, the energy $\omega_q$ is related via the dispersion relation $\omega_{q} = v_{s} q$ with $v_{s} = \sqrt{\gamma/\sigma} = 6.64\times 10^{3}$m/s (velocity of sound in graphene), $\sigma=$ mass density of the membrane and the Debye frequency of graphene $\omega_{D} = 65$ meV \cite{michel,LJ2011,DPC2013,Sengupta2016}. 

Let us now describe the Hamiltonian representing the atom-phonon interaction $H_{i} =$ $H_{bi}$ + $H_{ki}$. The atom-phonon interaction in the continuum is given by,
\begin{equation}\label{Hki}
H_{ki} = -\tilde{g}_{kb}\xi(c_{k}^{\dagger}b + b^{\dagger}c_{k})\sum_{q}(a_{q}+a_{q}^{\dagger}) 
\end{equation}
with $\tilde{g}_{kb}$ being the corresponding atom-phonon coupling in the continuum channel, $\xi$ is a frequency independent parameter that depends on the specific form of atom-excitation coupling and is given as $\xi = \sqrt{\hbar/(4L\sigma v_{s})}$\cite{DPC2013,Sengupta2016}. For the bound channel, we have:
\begin{equation}\label{Hbi}
H_{bi}=- \tilde{g}_{bb}\xi b^{\dagger}b\sum_{q}(a_{q}+a_{q}^{\dagger})
\end{equation}
with $\tilde{g}_{bb}$ as the atom-phonon coupling for an atom bound to the membrane \cite{DPC2013,Sengupta2016,Sengupta2017,clougherty2017}. For a surface without corrugations, a detailed procedure to calculate the atom-phonon couplings $\tilde{g}$'s from the Hamiltonian of the atom-surface scattering is given in Ref.~[\onlinecite{flatte}], this method is then suitably extended for a graphene membrane-atom interaction in Ref.~[\onlinecite{DPC2013}]. The general idea behind the procedure is to calculate the coupling constants through surface distortions/displacements, quantitatively relating it to the matrix element of the first derivative of the surface potential. The interaction between the atom and surface is modelled via the vdW potential $V_{0}(z)$ which is then Taylor expanded in small phonon-displacements in the target bath (a procedure valid for low-energies and temperatures) \cite{flatte,DPC2013}. For the vdW potential $V_{0}(z)$, one considers the asymptotic form of the long-range attractive interaction between the static flat graphene membrane and the neutral atom (equivalent to the Casimir-Polder potential between a 2D insulating solid and a neutral atom) given by,\cite{rubio, dion, DPC2013}
\begin{equation}\label{longrangevdw}
V_{0}(z) = -\frac{\pi C_{6}}{2}\bigg\{\frac{1}{z^{4}} - \frac{1}{(z^{2}+L^{2})^{2}}\bigg\}
\end{equation}
where $z$ is the distance between the particle and the membrane. Parameters like the physisorption potential and atom-phonon couplings for our model is then obtained by asymptotically treating the interaction potential leading to expressions for $\tilde{g}_{kb}  =  \langle k|V_{0}^{'}(z)|b\rangle$ and $\tilde{g}_{bb}  =  \langle b|V_{0}^{'}(z)|b\rangle$, where $|k\rangle$ and $|b\rangle$ represent the asymptotic continuum and bound state wave functions for the neutral atom\cite{flatte,DPC2013,dpc10}. For sufficiently low-energy incident atoms, the coupling $\tilde{g}_{kb}$ has a strong dependence on the incident energy of the incoming atom, such that $\tilde{g}_{kb}\propto \sqrt{E_{k}}$ \cite{DPC2013}. However, $\tilde{g}_{bb}$ is independent of the incident energy  and is much larger in magnitude than $\tilde{g}_{kb}$\cite{DPC2013}.

To derive the adsorption rate of the atom we do the following: we treat the atom-phonon coupling as perturbation and derive adsorption rates within a self-energy formalism using Green's functions \cite{Sengupta2016,Sengupta2017, sengupta2018}. We find that the terms in the perturbative expansion of atom self-energy are infrared divergent due to the contribution from the emission of low-energy phonons of the graphene membrane which we refer as the \textit{soft phonon} contribution \cite{Sengupta2016,Sengupta2017}. The problem of IR divergence is more pronounced at finite temperature because of the enhanced emission of thermal phonons due to the appearance of the Bose-Einstein function\cite{weldon91,Sengupta2016}.
Therefore, a crucial component for the derivation of  adsorption rates is to devise mathematical techniques to address these IR divergences and provide a well-unified theory that describes the role played by the emission of soft-phonons in the adsorption phenomenon.

\begin{figure}
    \centering
    \includegraphics[width=\columnwidth]{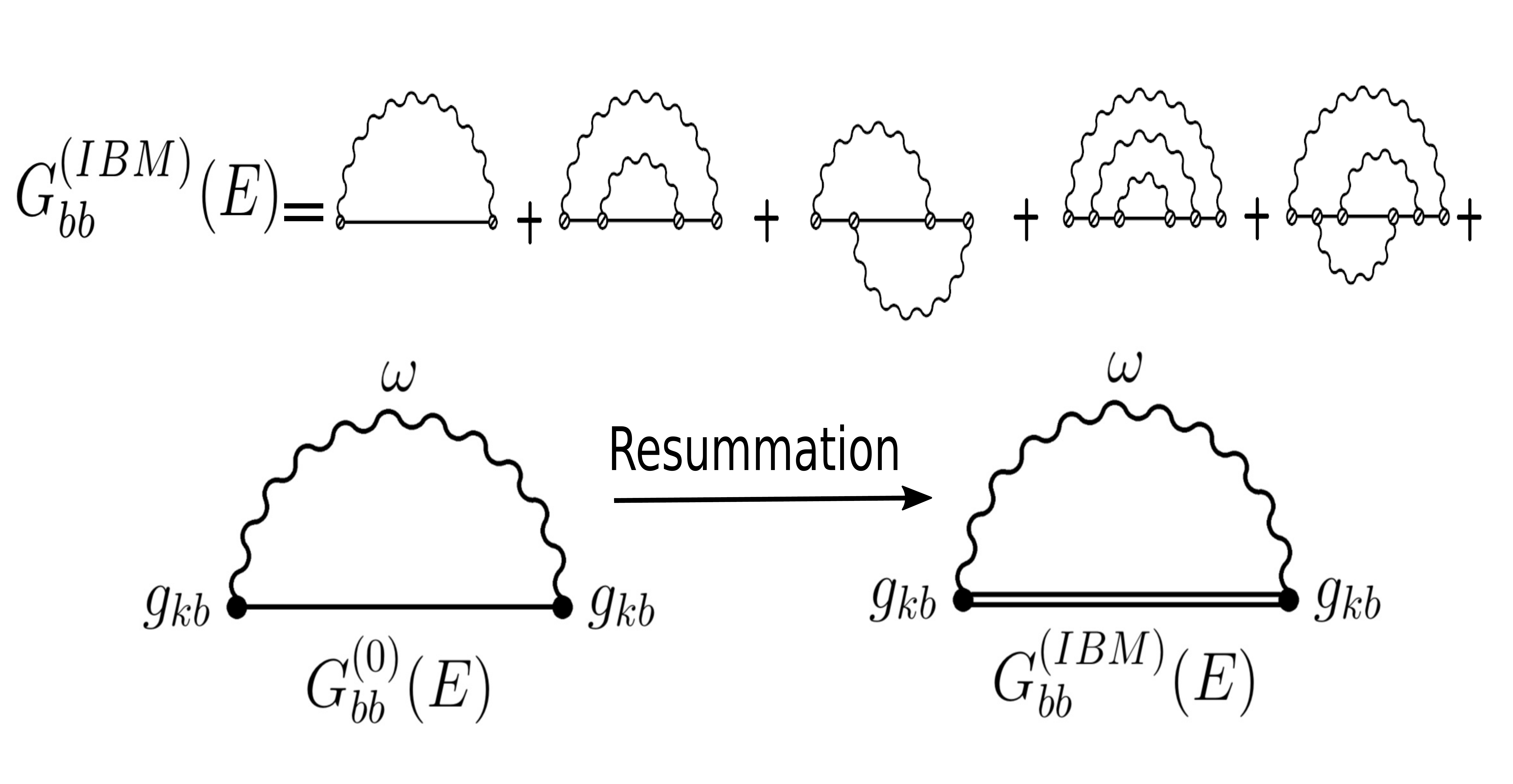}
    \caption{Top: Feynman diagrams corresponding to the exact bound state propagator $G_{bb}^{(IBM)}$ for the Independent Boson Model (IBM). As we see, $G_{bb}^{(IBM)}$ consists of a sum of Feynman diagrams of all types (vertex and rainbow) to all orders in pertubation in atom-phonon coupling $\tilde{g}_{bb}^{2}\xi^{2}$ (denoted by open dot in the self-energy diagrams) for transitions in the bound state $|b\rangle$. Bottom: 1 loop atom self-energy corresponding to adsorption mediated by a 1 phonon process (wiggly line) for transition from $|k\rangle \rightarrow |k\rangle$ via the bound state $|b\rangle$ (straight line denoting the bound state propagator $G_{bb}^{(0)}$). Under our resummation technique, we replace the bare $G_{bb}^{(0)}$ by the IBM propagator $G_{bb}^{(IBM)}$ (denoted by double lines) which represents the fully $dressed$ propagator with pertubations to all orders in atom-phonon coupling ($\tilde{g}_{bb}^{2}\xi^{2}$) in the bound. }
    \label{fig:FDIBM}
\end{figure}

We have devised such a method for addressing these IR divergences and predict adsorption rates. Our method displays a \textit{resummed} atom self-energy $\Sigma_{kk}$ which uses a \textit{fully dressed} bound state propagator (see Fig.~\ref{fig:FDIBM} for a general idea of this method) based on the exact solution of the independent boson model (IBM) \cite{Sengupta2016,Sengupta2017}. A natural question at this point would be: is it justified to use the IBM propagator to describe the physics of adsorption in our model? To the best of our knowledge, it seems it is imperative to provide a non-perturbative treatment for the bound channel compared to the continuum for the following reasons: (i) when treated perturbatively, the inclusion of the effects from the atom-phonon coupling in the bound state leads to severe IR divergences in the higher order self-energy terms\cite{Sengupta2016,Sengupta2017}, (ii) vertex renormalization results indicate an increase in the $\Gamma_{bb}$ vertex in the infrared limit \cite{Sengupta2016} and (iii) most importantly, our model satisfies the condition $\tilde{g}_{kb}\ll \tilde{g}_{bb}$ \cite{Sengupta2016,DPC2013,Sengupta2017}. Armed with this knowledge, let us now present the general procedure for the derivation of the many-body adsorption rate $\Gamma$.

Invoking Feynman rules for our model \cite{Sengupta2016,Sengupta2017} we will write the 1-loop atom self-energy $\Sigma_{kk}$ as\cite{Sengupta2016}
\begin{equation}\label{SKKbare}
\begin{split}
    \Sigma_{kk}(E)& = g_{kb}^{2}\sum_{q}\bigg[(n_{q}+1)G_{bb}^{(0)}(E-\omega_{q})\\
    &\quad+n_{q}G_{bb}^{(0)}(E+\omega_{q}) \bigg]\\
\end{split}
\end{equation}
where we introduced a label $g_{kb} = \tilde{g}_{kb}\xi$. The phonon occupation number with Bose-Einstein distribution function is given as $n_{q} =1/(e^{\omega_{q}/T}-1)$. The bare bound state propagator\cite{Sengupta2017} is written as $G_{bb}^{(0)} (E) = 1/(E+E_{b}+i\eta)$, $\eta\rightarrow 0^{+}$. 
Next we calculate the \textit{resummed} atom self-energy $\Sigma_{kk}^{(IBM)}$ by replacing the bare bound state propagator $G_{bb}^{(0)}$ by the \textit{fully dressed} $G_{bb}^{(IBM)}$ which is derived within the scheme of the independent boson model (see Fig.~\ref{fig:FDIBM}) \cite{Sengupta2016},
\begin{equation}\label{SKKIBM}
\begin{split}
    \Sigma_{kk}^{(IBM)} (E) &= g_{kb}^{2} \sum_{q}\bigg[(n_{q} +1) G_{bb}^{(IBM)} (E-\omega_{q})\\
     &\quad + n_{q} G_{bb}^{(IBM)} (E+\omega_{q})\bigg].\\
\end{split}
\end{equation}
We will write in details the properties and various features of the low temperature $G_{bb}^{(IBM)}$ in Sec.~\ref{sec:IBM}. Finally we derive the many-body adsorption rate $\Gamma$ using the imaginary part of $\Sigma_{kk}^{(IBM)}$ \cite{Sengupta2016,Sengupta2017}:
\begin{equation}\label{TRIBM}
    \Gamma = -2Z(E_{k})\mathcal{I}m \Sigma_{kk}^{(IBM)}(E_{k})
\end{equation}
where, $\mathcal{I}$m $\Sigma_{kk}^{(IBM)}$ gives the imaginary part of $\Sigma_{kk}^{(IBM)}$ and $\mathcal{Z}$ is the quasiparticle weight related to the real part of the self-energy $\Sigma_{kk}^{(IBM)}$,
\begin{equation}\label{RF}
    \mathcal{Z}(E) = \bigg[1-\bigg(\frac{\partial Re\Sigma_{kk}(E)}{\partial E}\bigg)\bigg]^{-1}.
\end{equation}

Before we proceed to the next section, let us present the result for the adsorption rate within the conventional 1$^{st}$ order perturbation theory i.e Fermi's golden rule \cite{Sengupta2016},
\begin{equation}\label{GR}
\Gamma_{0} = 2\pi g_{kb}^{2}\rho_{0}\bigg\{\frac{1}{\exp\bigg(\frac{E_{k}+E_{b}}{T}\bigg) -1}+1\bigg\}.
\end{equation}
where $\rho_{0}$ is the constant density of axisymmetric vibrational states\cite{Sengupta2016,Sengupta2017,clougherty2017}. Let us also introduce the label $g_{k}^{2} = g_{kb}^{2}\rho_{0}$. For sufficiently low-energy atomic hydrogen impinging on graphene membranes, $g_{k}^{2}$ is given by 0.7 $\mu$eV - 1.5 $\mu$eV\cite{Sengupta2016,Sengupta2017,DPC2013,clougherty2017}. We notice $\Gamma_{0}$ is independent of (i) atom-phonon coupling strength $\tilde{g}_{bb}$ and (ii) soft-phonon contribution ($\epsilon$).
As a test of our resummation technique and other comparative purposes related to the adsorption rate, we will use the golden rule result in Sec.~\ref{sec:MBA}. In the following sections, our aim is to understand the role played by the environment of thermal phonons towards the incoming atom in the bound state by investigating in detail the various features of the bound state propagator within the IBM.

\section{Low Temperature Bound State Propagator within the Independent Boson Model (IBM)}\label{sec:IBM}
In our model Hamiltonian (given by Eqs.~\ref{H0},~\ref{Hki} and \ref{Hbi}) if we focus only on the interaction of the atom and phonons in the bound state $|b\rangle$ and drop the terms that correspond to the continuum $|k\rangle$, we end up with the Hamiltonian of the Independent Boson Model given by,
\begin{equation}\label{IBMA}
H_{IBM} = -E_{b}b^{\dagger}b + \sum_{q}\omega_{q}a_{q}^{\dagger}a_{q} - g_{bb}b^{\dagger}b\sum_{q}(a_{q}+a_{q}^{\dagger}).
\end{equation}
where we have introduced the label $g_{bb} = \tilde{g}_{bb}\xi$.
Our aim in this section would be to re-write the exact solution of the IBM Hamiltonian in a suitable manner to incorporate the physics of the adsorption phenomenon.

With this aim in mind, let us begin with a canonical transformation using an operator $s = -g_{bb}b^{\dagger}b\sum_{q}(a_{q}^{\dagger}-a_{q})/\omega_{q}$ and apply it to each of the operators in Eq.~\ref{IBMA}. Using the Baker-Hausdorff lemma \cite {mahan}: $\bar{A} = e^{s}Ae^{-s} = A + [s,A] + (1/2!)[s,[s,A]]+\cdot \cdot$, we evaluate the new phonon operators:
\begin{equation}\label{an}
\bar{a_{q}} = e^{s}a_{q}e^{-s} = a_{q} + \frac{g_{bb}}{\omega_{q}}b^{\dagger}b \ ,
\end{equation}
\begin{equation}\label{and}
\bar{a_{q}}^{\dagger} = e^{s}a_{q}^{\dagger}e^{-s} = a_{q}^{\dagger} + \frac{g_{bb}}{\omega_{q}}b^{\dagger}b.
\end{equation}
here, we have used $[s,a_{q}] = (g_{bb}/\omega_{q}) b^{\dagger}b$. We see that under the canonical transformation, the phonon operators $a_{q}$ and $a_{q}^{\dagger}$ are displaced by an amount $(g_{bb}/\omega_{q})b^{\dagger}b$ to a new equilibrium position around which they vibrate with the initial frequency $\omega_{q}$\cite{nazarov_blanter_2009}. Quite naturally we ask: what is the physical reason behind such a displacement of the phonon fields? It seems that the presence of the adatom in the membrane leads to the polarization of the surface of the membrane which shifts the oscillators. This is quite similar to the case of the charged oscillator under a uniform electric field. The presence of the electric field causes the charge to displace to a new equilibrium position, around which it fluctuates with the same frequency as before \cite{mahan}. Now, let us write the new operator corresponding to the particle in the bound state:
\begin{equation}\label{tildeb}
\begin{split}
\bar{b} &= e^{s}be^{-s} = b\bigg(1+\sum_{q}\frac{g_{bb}}{\omega_{q}}(a_{q}^{\dagger}-a_{q})+\cdot\cdot\bigg)\\
&\quad = bX,
\end{split}
\end{equation}
here, we have used $[s,b] = \sum_{q}g_{bb}(a_{q}^{\dagger}-a_{q})b/\omega_{q}$ and introduced an operator $X$,
\begin{equation}
X = \exp\bigg(\sum_{q}\frac{g_{bb}}{\omega_{q}}(a_{q}^{\dagger}-a_{q})\bigg) 
\end{equation}
which we shall refer to as the coherent phonon bath displacement operator \cite{rae,royprx,royprb}. Indeed, we will see that $X$ is crucial to explain the fluctuations of the phonons around the atom. And finally under this canonical transformation $H_{IBM}$ modifies to:
\begin{equation}\label{nHam}
\begin{split}
\bar{H}_{IBM} &= e^{s}H_{IBM}e^{-s}\\
&\quad = - E_{b}\bar{b}^{\dagger}\bar{b}-g_{bb}b^{\dagger}b\sum_{q}\bigg(a_{q}+a_{q}^{\dagger} + \frac{2g_{bb}}{\omega_{q}}b^{\dagger}b\bigg)\\
&\quad +\sum_{q}\omega_{q}\bigg(a_{q}^{\dagger}+\frac{g_{bb}}{\omega_{q}}b^{\dagger}b\bigg)\bigg(a_{q}+\frac{g_{bb}}{\omega_{q}}b^{\dagger}b\bigg)\\
&\quad =  - (E_{b} + \Delta)b^{\dagger}b + \sum_{q}\omega_{q}a_{q}^{\dagger}a_{q}
\end{split}
\end{equation}
where, we have used $[X,b] = 0$ and $X^{\dagger} = X^{-1}$ which implies $\bar{b}^{\dagger}\bar{b} = b^{\dagger}b$ \cite{mahan}. The factor $\Delta$ is defined as the \textit{acoustic polaron} shift which appears as a result of the displacement of the phonon fields due to the presence of the atom in the phonon bath,
\begin{equation}\label{polaron}
\Delta = \sum_{q}\frac{g_{bb}^{2}}{\omega_{q}}
\end{equation} 
In the continuum limit $\sum_{q}\rightarrow \int_{\epsilon}^{\omega_{D}}\rho_{0}\mathrm{d}\omega$, the above Eq. reduces to:
\begin{equation}\label{contpolaron}
\Delta = g_{b}^{2}\int_{\epsilon}^{\omega_{D}}\frac{\mathrm{d}\omega}{\omega}
\end{equation}

Here, we have introduced the label $g_{bb}^{2}\rho_{0} = g_{b}^{2}$ and we stick to this notation for the rest of the paper. For atomic hydrogen impinging on graphene membranes $g_{b}^{2} = 0.06$ meV\cite{DPC2013,Sengupta2016,Sengupta2017,clougherty2017}.

A solution to Eq.~\ref{nHam} is realized within the IBM \cite{mahan} and is written with modifications pertaining to our model of adsorption as,
\begin{equation} \label{IBMP}
G_{bb}^{(IBM)}(t) = -ie^{-it(-E_{b}-\Delta)} e^{-\tilde{\phi}(t)}.
\end{equation}
We notice that the canonical transformation has led to a propagator in which the contributions due to the atom and phonon terms are well separated. Let us first look at the phonon contribution which is related to the thermal average over the phonon modes leading to the phonon bath correlator \cite{mahan,hohenester2010,rc1,rc2}
\begin{equation}\label{phononcorr}
\langle X(t) X^{\dagger}(0)\rangle = \exp[-\tilde{\phi}(t)]
\end{equation}
with the IBM phase factor\cite{mahan,Sengupta2016}
\begin{equation}\label{phi}
\tilde{\phi}(t) = \sum_{q}\bigg(\frac{g_{bb}}{\omega_{q}}\bigg)^{2}\bigg[n_{q}(1-e^{i\omega_{q}t}) + (n_{q}+1)(1-e^{-i\omega_{q}t})\bigg]
\end{equation}
which can be further decomposed into
\begin{equation}\label{decphi}
\tilde{\phi}(t) = \phi(0) + \phi(t).
\end{equation}
Within the continuum limit, we define the above terms as:
\begin{equation}\label{phi0}
    \phi(0) = \int_{\epsilon}^{\omega_{D}}\frac{g_{b}^{2}}{\omega^{2}}\bigg[2n+1\bigg]\mathrm{d}\omega
\end{equation}
\begin{equation}\label{phit}
    \phi(t) = \int_{\epsilon}^{\omega_{D}}\frac{g_{b}^{2}}{\omega^{2}}\bigg[(n+1)e^{-i\omega t} + ne^{i\omega t}\bigg]\mathrm{d}\omega .
\end{equation}
where $n =1/(e^{\omega/T}-1)$. Plugging Eq.~\ref{phi0} and \ref{phit} in Eq.~\ref{IBMP}, we derive the $\textit{fully dressed}$ bound state propagator $G_{bb}^{(IBM)}(t)$:
\begin{equation}\label{prop}
G_{bb}^{(IBM)} (t)= -i \exp\bigg\{it(E_{b}+\Lambda) \bigg\}\exp\bigg[-\phi(0)+\mathcal{R}e \  \phi(t)\bigg].
\end{equation}
Here, $\Lambda$ $\textit{renormalizes}$ the energy of the bound atom and includes the contribution from two terms: (i) the acoustic polaron shift $\Delta$ and (ii) imaginary part of the phonon bath correlator $\mathcal{I}m \ \phi(t)$ such that:
\begin{equation}\label{renorm}
\begin{split}
\Lambda & = \Delta + \frac{\mathcal{I}m \ \phi(t)}{t}\\
&\quad = g_{b}^{2}\int_{\epsilon}^{\omega_{D}}\frac{\mathrm{d}\omega}{\omega} - g_{b}^{2}\int_{\epsilon}^{\omega_{D}}\frac{\sin(\omega t)\mathrm{d}\omega}{\omega^{2}t}.\\
\end{split}
\end{equation}
The decay of the propagator $G_{bb}^{(IBM)}(t)$ on the other hand, is given by
\begin{equation}\label{decay}
\mathcal{S}=\exp\bigg[-\phi(0)+\mathcal{R}e \  \phi(t)\bigg] \ ,
\end{equation}
which comprises the real part of $\phi(t)$:
\begin{equation}\label{realpc}
\mathcal{R}e \ \phi(t) = g_{b}^{2}\int_{\epsilon}^{\omega_{D}}\bigg[\frac{(2n+1)\cos(\omega t)}{\omega^{2}}\bigg]\mathrm{d}\omega
\end{equation}
and a shift $\exp[-\phi(0)]$ linked to the Franck-Condon factor. This in principle is related to the phonon contribution by the relation, $\langle X\rangle = \exp[-\phi(0)/2]$\cite{rae,mahan,rc1}.

For all purposes related to the calculation of the many-body adsorption rate, we require the Fourier transform of Eq.~\ref{prop}. We write them as the following:
\begin{equation}\label{FSrealG}
\begin{split}
\mathcal{R}e G_{bb}(E+E_{b}) & =\int_{0}^{\infty}\mathrm{d}t \sin\bigg\{t(E+E_{b}+\Lambda) \bigg\}\\
&\quad \times \exp\bigg[-\phi(0)+\mathcal{R}e \  \phi(t)\bigg]\\
\end{split}
\end{equation}
and
\begin{equation}\label{FSimagG}
\begin{split}
\-|\mathcal{I}m G_{bb}(E+E_{b})\-|  & =\int_{0}^{\infty}\mathrm{d}t \cos\bigg\{t(E+E_{b}+\Lambda) \bigg\}\\
&\quad \times \exp\bigg[-\phi(0)+\mathcal{R}e \  \phi(t)\bigg].\\
\end{split}
\end{equation}
 
Before we calculate the resummed atom self-energy $\Sigma_{kk}$ using Eqs.~\ref{FSrealG} and \ref{FSimagG}, let us study the effect of the phonon bath on the renormalization and decay of the bound atom with a special focus on the time evolution, temperature $T$ and coupling strength $g_{b}^{2}$ . We envision these effects to alter the response of the phonons towards the adsorption phenomenon. Our next section will be devoted to this.

\section{Effects of the phonon correlator on the renormalization and decay of the bound state propagator}
\label{sec:phononcorr}
We devote this section to investigate the effects of time, temperature and coupling on the phonon bath correlator and also the renormalization parameter. It seems to us that the first step to accomplish this would be to define an effective $parameter$ which would accommodate the effects of temperature $T$ and coupling $g_{b}^{2}$ in it. 
In principle, we accomplish this by setting up a transformation of variable which would naturally give rise to such a parameter. Let us define the transformation of variable,

\begin{equation}\label{trans}
\tilde{\omega} = \frac{\omega}{\sqrt{g_{b}^{2}T}}
\end{equation}
which leads to the following dimensionless parameters for the low-energy $\textit{infrared}$ scale $\epsilon$ and maximum Debye frequency $\omega_{D}$,
\begin{equation}\label{limits}
\epsilon\rightarrow \tilde{\epsilon} = \frac{\epsilon}{\sqrt{g_{b}^{2}T}} , \ \omega_{D} \rightarrow \tilde{\omega}_{D} = \frac{\omega_{D}}{\sqrt{g_{b}^{2}T}}.
\end{equation}
Also as result of the transformation, we derive a new characteristic dimensionless time scale that comprises the effects of time t:
\begin{equation}\label{time}
\tau = \bigg(\sqrt{g_{b}^{2}T}\bigg) t
\end{equation}

As we see, the parameter $\tilde{\epsilon}$ contains in it the soft phonon energy scale $\epsilon$, length of the membrane $L$ (since, $\epsilon=v_{s}/L)$, coupling $g_{b}^{2}$ and the effects of temperature $T$. Thus we study the time-dependence of the decay and renormalization of the bound state propagator as a function of $\tilde{\epsilon}$. We will perform the study with respect to the dimensionless time scale $\tau$. The effectiveness of such a choice will be clear to us shortly. In the following subsections we apply the transformations given by Eqs.~\ref{trans}, ~\ref{limits} and ~\ref{time} to the decay and the renormalization factors.
\subsection{Decay of the propagator as a function of $\tilde {\epsilon}$}
\label{sec:phonondecay}
\begin{figure}
    \centering
    \includegraphics[width=\columnwidth]{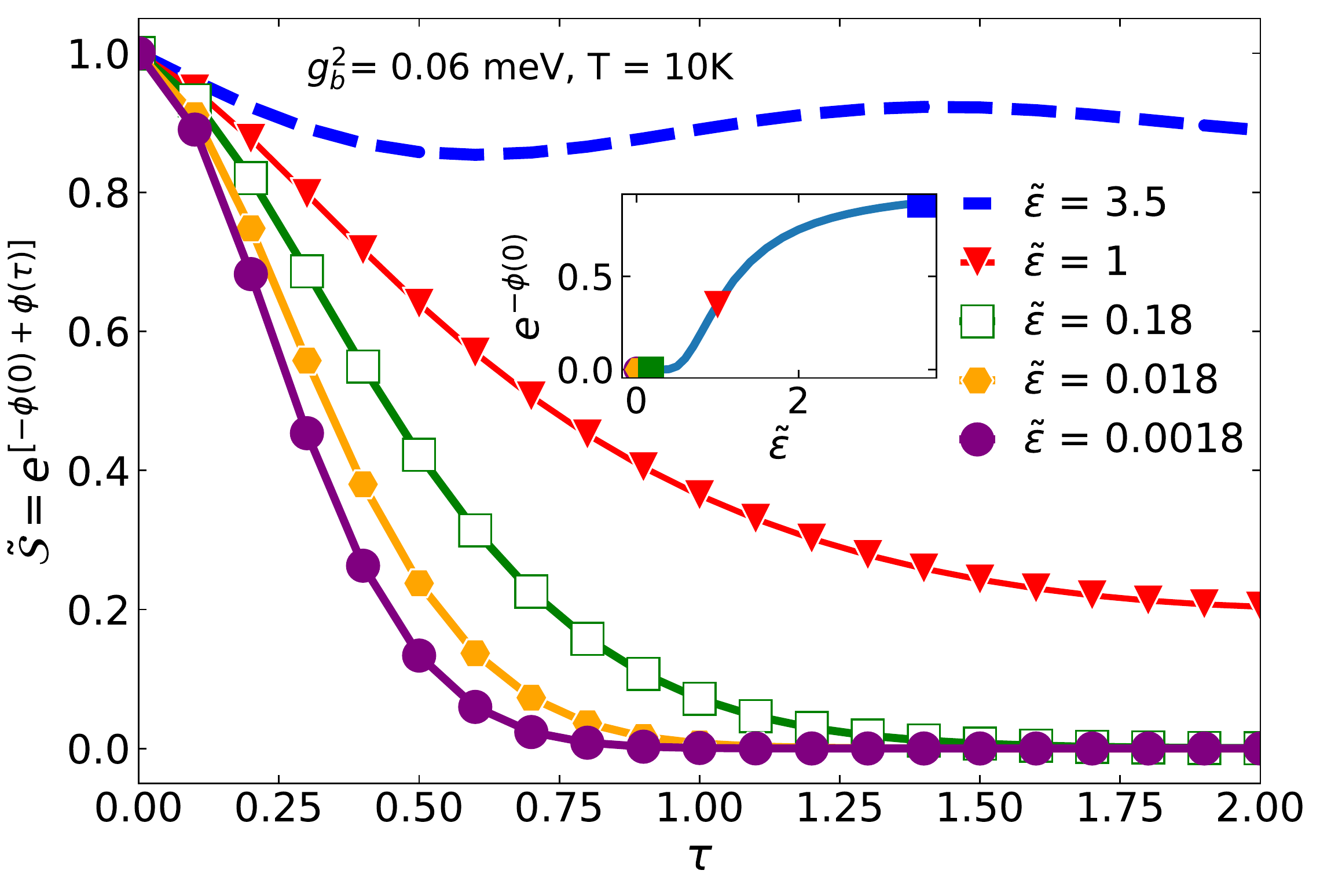}
    \caption{For graphene membranes with coupling $g_{b}^{2} = 0.06$ meV\cite{DPC2013,clougherty2017,Sengupta2016,Sengupta2017} maintained at $T= 10$ K = 0.862 meV, we plot the effect of the dimensionless $\textit{infrared}$ scale $\tilde{\epsilon}$ on the decay factor $\mathcal{\tilde{S}}$ as a function of the dimensionless characteristic time scale $\tau =(\sqrt{g_{b}^{2}T})t$. The range of membrane sizes used are $L = 5$ nm, 20 nm, 100 nm, 1$\mu$m and 10$\mu$m which correspond to $\tilde{\epsilon} =$ 3.5, 1, 0.18, 0.018 and 0.0018 respectively. While for smaller values of $\tilde{\epsilon}$ ($\tilde{\epsilon}<1$), we notice a rapid loss of phonon coherence within $\tau<1$, higher values of $\tilde{\epsilon}$ ($\tilde{\epsilon} >1$) retain their coherence and slowly saturate to non-zero residual values for $\tau>1$. We relate this long-time residue to the shift due to the Franck-Condon (FC) factor given by $\exp[-\phi(0)]$. The inset plot shows the corresponding variation of $\exp[-\phi(0)]$ with $\tilde{\epsilon}$. For $\tilde{\epsilon}\ll1$, the shift due to FC is generally zero and shows appreciable non-zero values only for $\tilde{\epsilon} \geq1$.}
    \label{fig:decayt}
\end{figure}

Under the chosen transformation of variables, the decay term $\mathcal{S}$ given by Eq.~\ref{decay} modifies to:
\begin{equation}\label{transde}
\begin{split}
\tilde{\mathcal{S}} &= \exp\bigg[-\int_{\tilde{\epsilon}}^{\tilde{\omega}_{D}}\sqrt{\frac{g_{b}^{2}}{T}}\frac{1}{\tilde{\omega}^{2}}\bigg\{\frac{2}{\exp(\tilde{\omega}\sqrt{g_{b}^{2}/T})-1}+1\bigg\}\\
&\quad\times\bigg\{1-\cos(\tilde{\omega}\tau) \bigg\}\mathrm{d}\tilde{\omega}\bigg]\\
\end{split}
\end{equation}
where we have used the transformed versions of Eqs.~\ref{phi0} and \ref{realpc}. We solve the integral under the parenthesis numerically for various values of $\tilde{\epsilon}$ as a function of the characteristic dimensionless time scale $\tau$ and plot the decay factor $\mathcal {\tilde{S}}$ vs $\tau$ in Fig.~\ref{fig:decayt}. 

As we have seen in the previous section, $\mathcal{\tilde{S}}$ represents the phonon bath correlator (given by Eq.~\ref{phononcorr}) and physically corresponds to the $\textit{dephasing}$ of the phonons. The general feature as seen in Fig.~\ref{fig:decayt} is a loss of coherence of the phonons following a power law decay. However, we observe two regimes: (i) for phonons corresponding to $\tilde{\epsilon} <1$, there is a rapid loss of coherence within $\tau <1$ and beyond $\tau=1$, phonons have completely lost their coherence such that $\tilde{S}\approx0$ and (ii) for phonons corresponding to $\tilde{\epsilon} >1$, there is a much slower loss of coherence and in fact beyond $\tau =1$ they do not show a complete decay but saturate to non-zero residual values. Let us ask what controls the phonon bath correlator function beyond $\tau =1$? While phonons lose their coherence with the evolution of time, they finally saturate to residual values that are represented by the time-independent Franck-Condon factor (FC) given by $\exp[-\phi(0)]$. We indeed observe a complete match of the long-time values of $\mathcal{\tilde{S}}$ with the FC factors calculated for the corresponding values of $\tilde{\epsilon}$ (see inset of Fig.~\ref{fig:decayt}).
Therefore for $\tau<1$, we observe short-time phonon dynamics that correspond to $\tilde{\epsilon}<1$ or more precisely $\epsilon<\sqrt{g_{b}^{2}T}$ and for $\tilde{\epsilon}>1$ i.e $\epsilon>\sqrt{g_{b}^{2}T}$ phonons exhibit long-time dynamics in regime $\tau>1$ .

\begin{figure}
    \centering
    \includegraphics[width=\columnwidth]{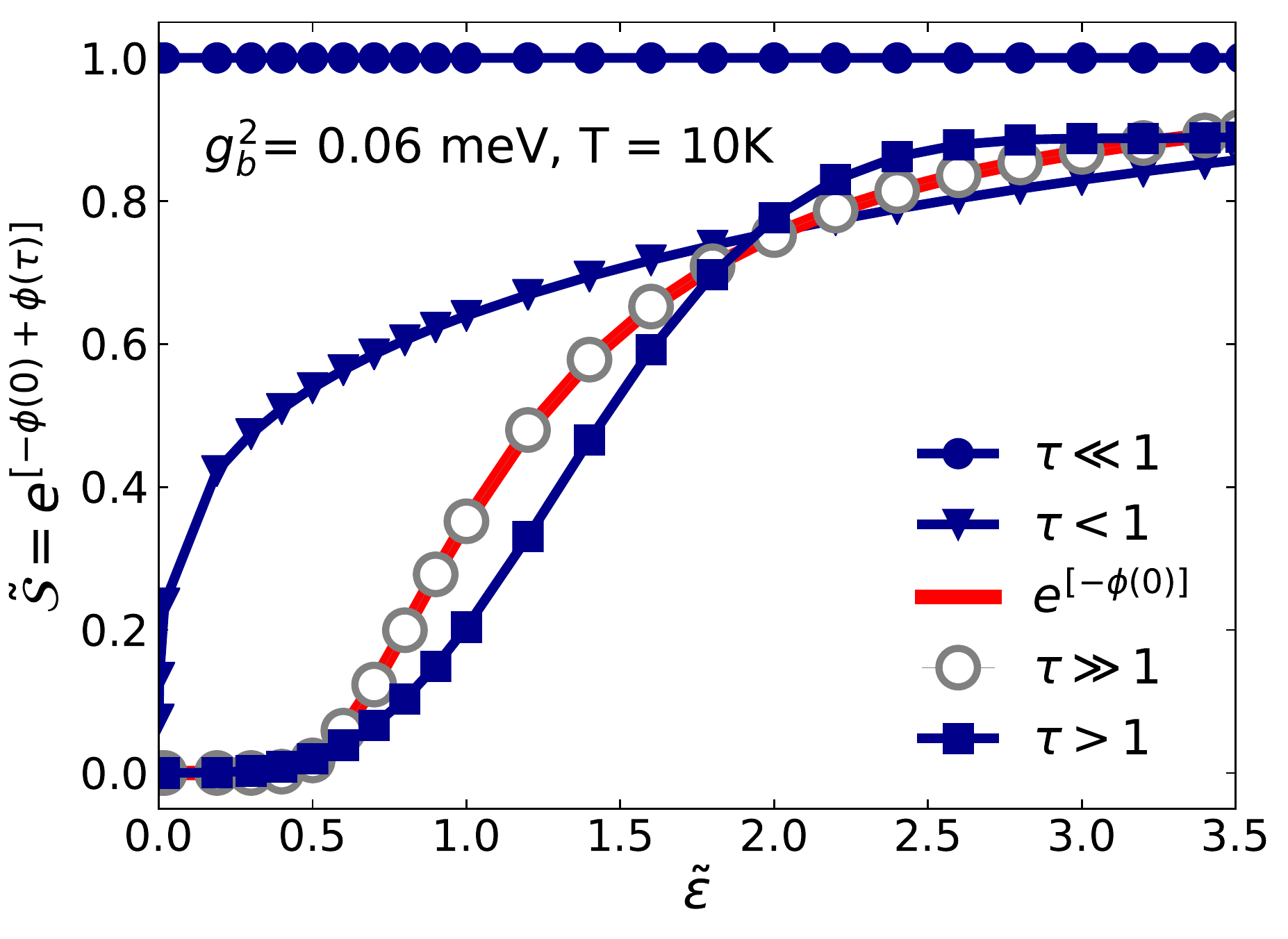}
    \caption{Dependence of the decay factor $\mathcal{\tilde{S}}$
    on $\tilde{\epsilon}$ for different regimes of the dimensionless characteristic time scale $\tau =(\sqrt{g_{b}^{2}T})t$. Maximum phonon coherence is observed for the regime $\tau\ll1$. With increasing $\tau$, phonon coherence is seen to decrease and beyond $\tau=1$, the shift due to Franck-Condon factor $\exp[-\phi(0)]$ starts to emerge. For phonons exhibiting short-time phonon dynamics ($\tilde{\epsilon}<0.5$) an absolute loss of phonon coherence beyond $\tau=1$ is observed. The long-time ($\tau\gg1$) response of $\mathcal{\tilde{S}}$ corresponds to the shift due to the Franck-Condon factor. }
    \label{fig:decaye}
\end{figure}

Variation of the decay $\mathcal{\tilde{S}}$ as a function of $\tilde{\epsilon}$ for different regimes of the dimensionless characteristic time scale $\tau =(\sqrt{g_{b}^{2}T})t$ is shown in Fig.~\ref{fig:decaye}. We observe that the function is within the bounds $0\leq\mathcal{\tilde{S}}\leq1$. The shift due to the Franck-Condon factor $\exp[-\phi(0)]$ matches exactly with the long-time ($\tau\gg1$) response of $\mathcal{\tilde{S}}$. For $\tau\ll1$, phonon coherence is maximum (also, see Fig.~\ref{fig:decayt}) and it is this time-dependent contribution that completely cancels out the one arising from the time-independent factor $\phi(0)$, hence there is no appreciable change in the decay function and $\mathcal{\tilde{S}} \sim 1$. However, with increasing $\tau$, phonon coherence starts to decay and the shift due to FC factor starts to emerge. For $\tau>1$ and $\tau\gg1$, $\mathcal{\tilde{S}} \approx 0$ for $\tilde{\epsilon}<0.5$. We relate this to the absolute loss of phonon coherence beyond $\tau=1$ for the phonons exhibiting the short-time phonon dynamics (see Fig.~\ref{fig:decayt}). The regime $\tau<1$ is interesting as it shows a logarithmic increase in $\mathcal{\tilde{S}}$ with increasing $\tilde{\epsilon}$. For $\tilde{\epsilon}>2$, $\mathcal{\tilde{S}}$ is finite and comparable for all regimes of $\tau$. We understand this behavior by relating this to the long-time phonon dynamics where the phonons in the regime $\tilde{\epsilon}>1$ retain their coherence for longer times.

We now ask if the renormalization factor $\Lambda$ exhibits similar dependence on $\tau$ and $\tilde{\epsilon}$ just as the decay $\mathcal{\tilde{S}}$.

\subsection{Renormalization of the bound state energy as a function of $\tilde{\epsilon}$}
\begin{figure}
    \centering
    \includegraphics[width=0.45\textwidth]{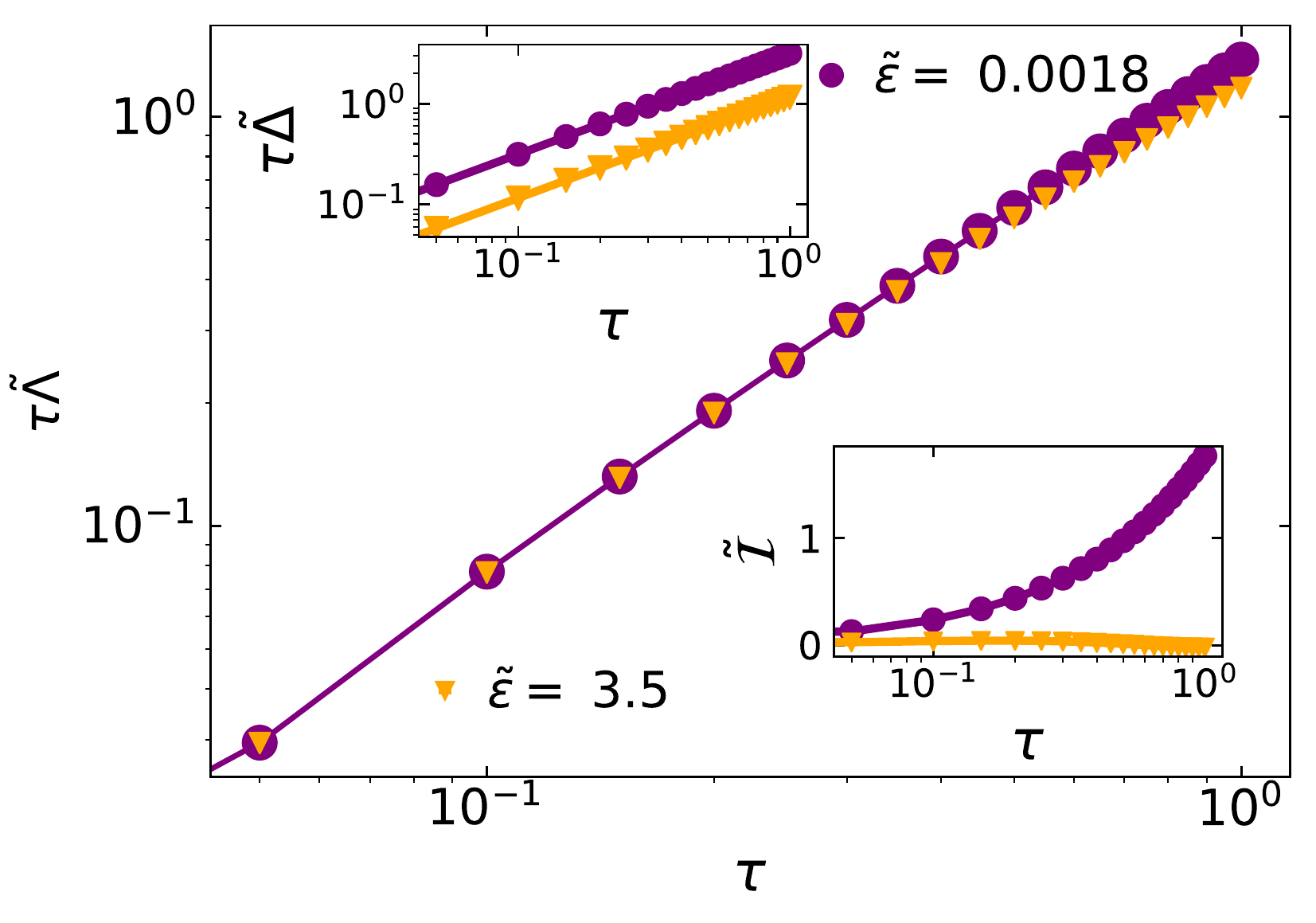}
    \includegraphics[width=0.44\textwidth]{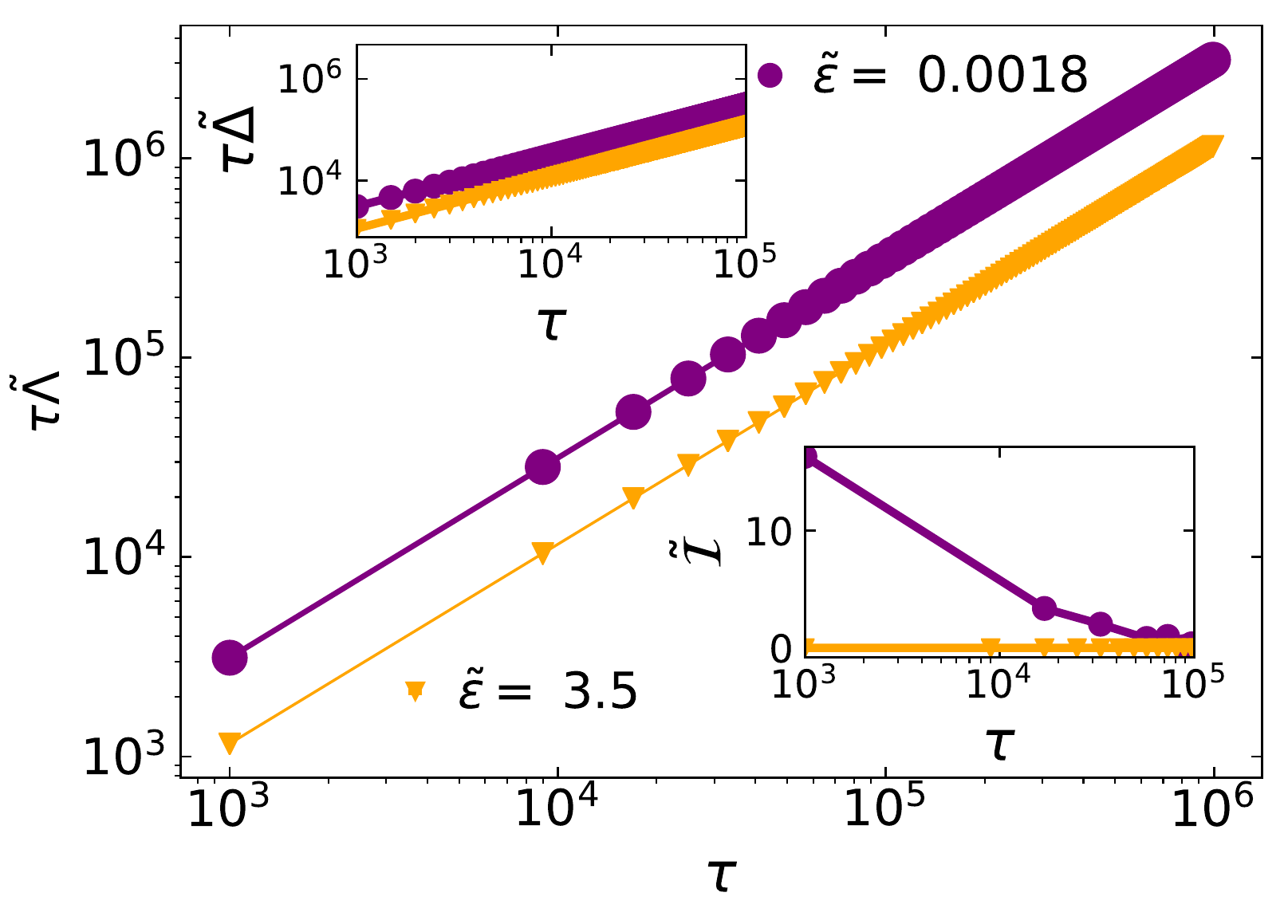}
    \caption{Variation of the renormalization factor $\tau\tilde{\Lambda}$ corresponding to two regimes of $\tau$ for various values of $\tilde{\epsilon}$. Negligible renormalization effects are observed in the short-time regime $\tau\ll1$ (top panel) with small contributions both from the polaron $\tau\tilde{\Delta}$ (left inset) and imaginary part of the phonon correlator  $\tilde{\mathcal{I}}$ (right inset). In the long-time regime $\tau\gg1$, massive renormalization effects are observed with large contribution from the polaronic term. }
    \label{fig:renormt}
\end{figure}
From Eq.~\ref{FSimagG}, we define a function $\mathcal{P}$ which encloses the renormalization factor $\Lambda$,
\begin{equation}\label{renormph}
\mathcal{P}=\cos\bigg\{t(E+E_{b}+\Lambda) \bigg\}.
\end{equation}
Under the chosen transformation, we write:
\begin{equation}\label{transrenorm}
\begin{split}
\tilde{\mathcal{P}} &= \cos\bigg[\tau\bigg\{\frac{E+E_{b}}{\sqrt{g_{b}^{2}T}}+\sqrt{\frac{g_{b}^{2}}{T}}\int_{\tilde{\epsilon}}^{\tilde{\omega}_{D}}\frac{\mathrm{d}\tilde{\omega}}{\tilde{\omega}} \\
&\quad - \sqrt{\frac{g_{b}^{2}}{T}}\int_{\tilde{\epsilon}}^{\tilde{\omega}_{D}}\frac{\sin(\tilde{\omega}\tau)}{\tilde{\omega}^{2}\tau}\mathrm{d}\tilde{\omega}\bigg\}\bigg]
\end{split}
\end{equation}
with the transformed renormalization term:
\begin{equation}\label{renormlam}
\tilde{\Lambda} = \bigg[\sqrt{\frac{g_{b}^{2}}{T}}\int_{\tilde{\epsilon}}^{\tilde{\omega}_{D}}\frac{\mathrm{d}\tilde{\omega}}{\tilde{\omega}} - \sqrt{\frac{g_{b}^{2}}{T}}\int_{\tilde{\epsilon}}^{\tilde{\omega}_{D}}\frac{\sin(\tilde{\omega}\tau)}{\tilde{\omega}^{2}\tau}\mathrm{d}\tilde{\omega}\bigg]
\end{equation}
As discussed before in Sec.~\ref{sec:IBM}, there is a competition between two physical effects that govern the renormalization $\tilde{\Lambda}$: (i) the acoustic polaronic shift $\tilde{\Delta}$ and (ii) the imaginary part of the phonon bath correlator $\mathcal{\tilde{I}}$ (given by the first and second terms of Eq.~\ref{renormlam} respectively). 

\begin{figure}
    \centering
    \includegraphics[width=\columnwidth]{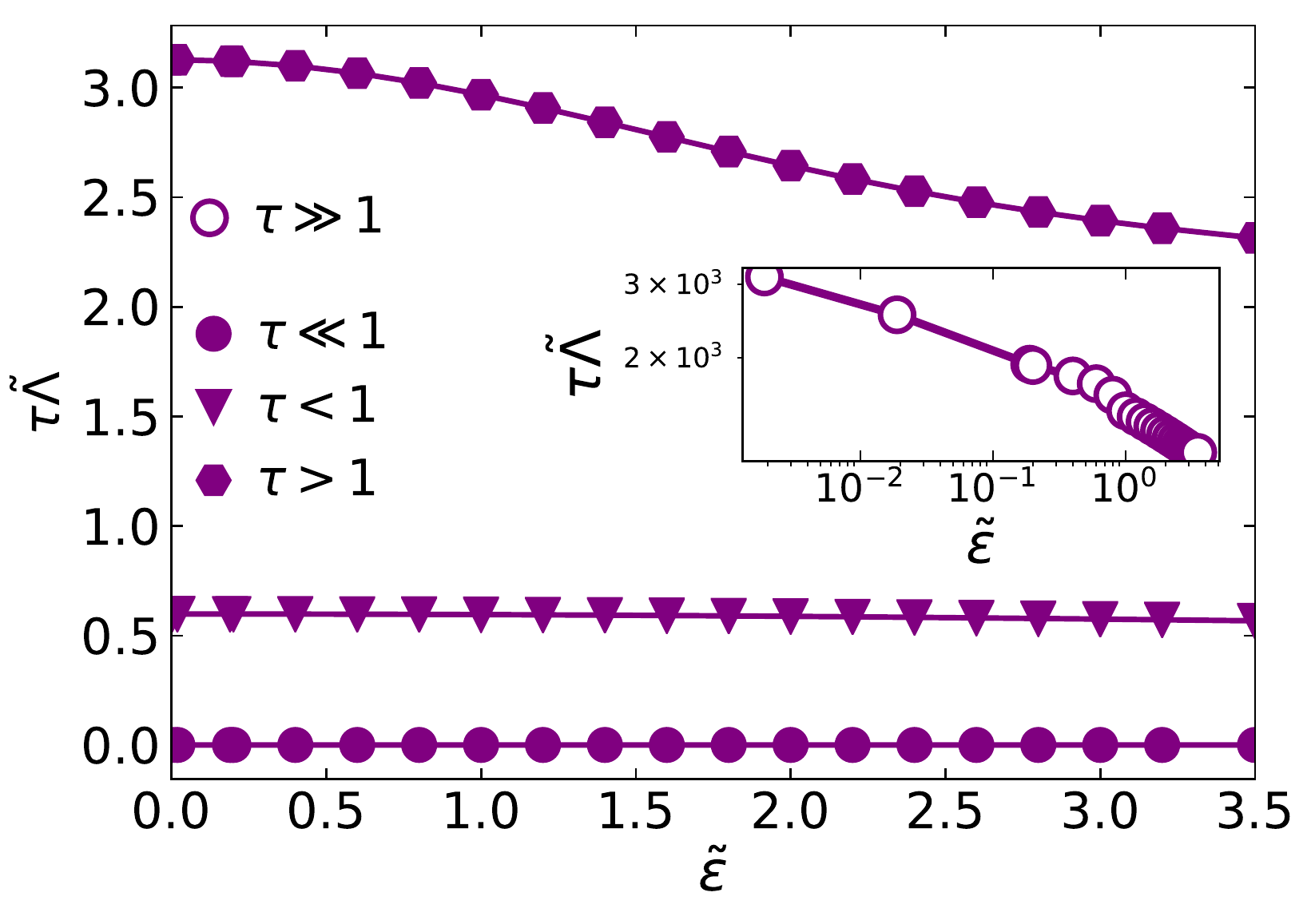}
    \caption{For the short time regime corresponding to $\tau<1$, small renormalization ($\tau\tilde{\Lambda}$) effect without any significant dependence on $\tilde{\epsilon}$ is observed. With increasing $\tau$, renormalization effect is seen to increase for the phonons satisfying $\tilde{\epsilon}<1$.}
    \label{fig:renorme}
\end{figure}

We plot the time dependence of the renormalization factor $\tau\tilde{\Lambda}$ for various values of the dimensionless IR cut-off $\tilde{\epsilon}$ for 2 regimes $0<\tau\leq1$ (top panel of Fig.~\ref{fig:renormt}) and $\tau\gg1$ (bottom panel of Fig.~\ref{fig:renormt}). For $0<\tau\leq1$ (top panel), we observe comparable and small contributions (compared to $E_{b}$) from the polaron (left inset) and the imaginary part of phonon correlator (right inset) which thus results in an overall small renormalization factor. We relate this to the fact that the phonons in this regime are undergoing $\textit{dephasing}$ and have not fully lost their coherence (as seen in the previous subsection). As a result, while phonons dephase with time, the polaron ($\tau\tilde{\Delta}$) starts to grow and hence is negligibly small at the onset of time. However, for the asymptotically large $\tau$ regime (bottom panel), we observe a huge contribution from the polaron (left inset) towards the renormalization compared to the imaginary part of the phonon bath correlator which gradually decays off (right inset). Once again, we relate this to the complete transfer of coherence from the phonon bath towards the formation of the polaron which now grows logarithmically without any decay. Variation of the renormalization $\tau\tilde{\Lambda}$ as a function of $\tilde{\epsilon}$ for various values of $\tau$ is shown in Fig.~\ref{fig:renorme}. For small values of $\tau$ ($\tau\ll1, \tau<1$), we observe negligible variation of the renormalization $\tau\tilde{\Lambda}$ with $\tilde{\epsilon}$. The modest renormalization factors can be related to the comparable and tiny contributions from the polaron and phonon bath correlator. However, the general trend shows an increase in $\tau\tilde{\Lambda}$ with increasing $\tau$ and the effect is seen to be more pronounced for $\tilde{\epsilon}<1$. The total loss of phonon coherence beyond $\tau=1$ for phonons exhibiting short-time phonon dynamics ($\tilde{\epsilon}<1$) can be attributed to the large renormalization factor. See inset plot of Fig.~\ref{fig:renorme} for the effect in the asymptotically large times ($\tau\gg1$).

In the next section, we will calculate the many-body adsorption rates $\Gamma$ following the procedure given in Sec.~\ref{sec:prelim} and see how the polaronic shifts and decay of the atom propagator due to phonons affect the adsorption rates.

\section{Many-body adsorption rate}\label{sec:MBA}
In this section, we will begin our analysis for suspended graphene membranes maintained at 10 K with sizes ranging from 100 nm $\sim$ 10 $\mu$m which correspond to $\tilde{\epsilon} = 0.18 \sim$ 0.0018 (for atom-phonon coupling in the bath $g_{b}^{2}$ =0.06 meV). As we have seen before in Sec.~\ref{sec:phononcorr}, there seems to be the existence of two regimes corresponding to the characteristic time scale $\tau$. While in $\tau\leq1$, phonons exhibit observable dynamics, in the other regime characterized by $\tau> 1$, phonon dynamics seem to completely die off with massive renormalization effects. In subsections \ref{sec:smalltime} and \ref{sec:longtimes}, we provide procedures which separately evaluate the contribution of the time regimes $0\leq\tau\leq1$ and $1<\tau\leq\infty$ to the adsorption rate. Finally in subsection~\ref{sec:fulltime}, we provide the result for the total many-body adsorption rate which is the sum of the contribution from regimes $0\leq\tau\leq1$ and $1<\tau\leq\infty$.

Let us begin with the expression for the imaginary part of the atom self-energy given by Eq.~\ref{SKKIBM}, re-written under the transformation of variables (given by Eqs.~\ref{trans},~\ref{limits} and \ref{time}),
\begin{equation}\label{IMSKK}
\begin{split}
\mathcal{I}m\Sigma_{kk}^{(IBM)} &= g_{k}^{2}\bigg[\int_{\tilde{\epsilon}}^{\tilde{\omega_{D}}}\mathrm{d}\tilde{\omega}\bigg\{\frac{1}{\exp(\tilde{\omega}\sqrt(g_{b}^{2}/T)-1}+1\bigg\}\\
&\quad\mathcal{I}m\tilde{G}_{bb}(\tilde{E}_{s}-\tilde{\omega})\bigg]\\
&\quad + g_{k}^{2}\bigg[\int_{\tilde{\epsilon}}^{\tilde{\omega_{D}}}\mathrm{d}\tilde{\omega}\bigg\{\frac{1}{\exp(\tilde{\omega}\sqrt(g_{b}^{2}/T)-1}\bigg\}\\
&\quad\mathcal{I}m\tilde{G}_{bb}(\tilde{E}_{s}+\tilde{\omega})\bigg]\\
\end{split}
\end{equation}
Here we have introduced the labels $\tilde{E}_{s}\equiv (E+E_{b})/\sqrt{g_{b}^{2}T}$.

The above equation can also be written as
\begin{equation}\label{spec}
    \mathcal{I}m \Sigma_{kk}^{(IBM)}= g_{k}^{2}\int_{\tilde{\epsilon}}^{\tilde{\omega_{D}}}\mathrm{d}\tilde{\omega}\mathcal{F} = g_{k}^{2}\int_{\tilde{\epsilon}}^{\tilde{\omega_{D}}}\mathrm{d}\tilde{\omega} \bigg[\mathcal{F}^{(em)} + \mathcal{F} ^{(abs)}\bigg] ,
\end{equation}
where $\mathcal{F}^{(em)}$ and $\mathcal{F}^{(abs)}$ correspond to processes for emission and absorption of phonons, respectively. We write these as
\begin{equation}\label{Fem}
\begin{split}
\mathcal{F}^{(em)} &= \bigg[\bigg\{\frac{1}{\exp(\tilde{\omega}\sqrt(g_{b}^{2}/T)-1}+1\bigg\}\\
&\quad\mathcal{I}m\tilde{G}_{bb}(\tilde{E}_{s}-\tilde{\omega})\bigg]\\
\end{split}
\end{equation}
and,
\begin{equation}\label{Fabs}
\begin{split}
    \mathcal{F}^{(abs)}&=\bigg[\bigg\{\frac{1}{\exp(\tilde{\omega}\sqrt(g_{b}^{2}/T)-1}\bigg\}\\
&\quad\mathcal{I}m\tilde{G}_{bb}(\tilde{E}_{s}+\tilde{\omega})\bigg]\\
\end{split}
\end{equation}

As a matter of fact, a careful inspection of Eqs.~\ref{IMSKK}, \ref{Fem} and \ref{Fabs} reveals that we can relate the functions $\mathcal{F}^{(em)}$ and $\mathcal{F}^{(abs)}$ to the spectral weights associated with the processes of emission and absorption of phonons since they are related to the imaginary part of the bound state propagator. With the above equations, let us now look at the contribution from the two different regimes of $\tau$. We begin with $0\leq\tau\leq 1$. 

\subsection{Contribution from regime $0\leq\tau\leq1$}
\label{sec:smalltime}

In this subsection we will calculate the contribution to $\Gamma$ from the regime $0\leq\tau\leq1$ by using equations Eqs.~\ref{IMSKK},~\ref{Fem} and \ref{Fabs}. We will plug the bound state propagator given by Eq.~\ref{FSimagG} with respective transformation of variables and integrated for $0\leq \tau \leq 1$. We start with a study on the dependence of the function $\mathcal{F}^{(em)}$ on the entire phonon frequency scale ($\tilde{\omega})$ for our chosen values of $\tilde{\epsilon}$.
\begin{figure}[t]
\begin{centering}
\includegraphics[width=\columnwidth]{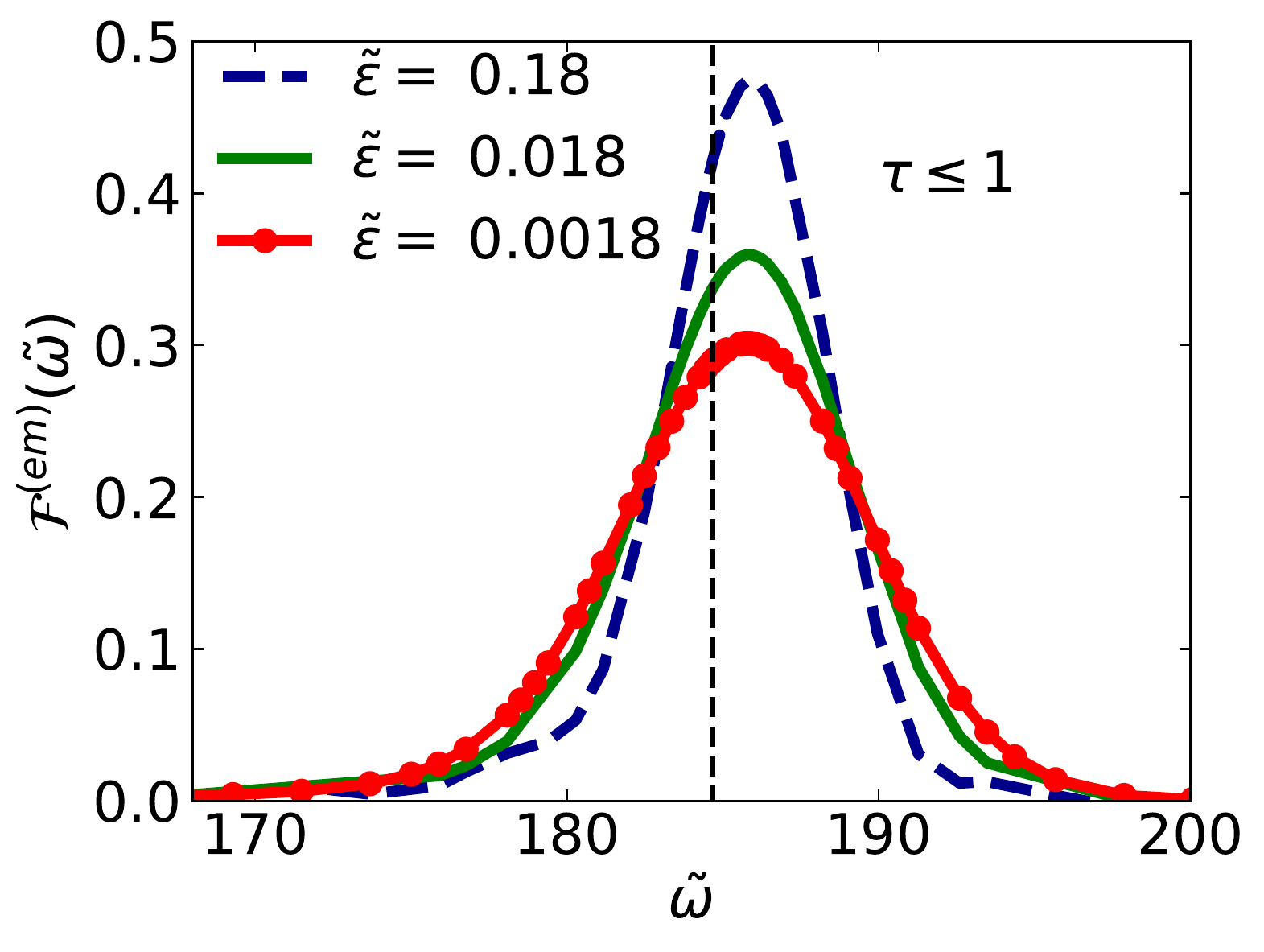}
\caption{ Variation of the spectral weight $\mathcal{F}^{(em)}$  as a function of $\tilde{\omega}$ for various values of $\tilde{\epsilon}$ in the regime $\tau\leq 1$. Phonon-broadened peaks are observed around the vertical black dashed line which corresponds to phonon frequency for the bound state energy $\tilde{\omega} = \tilde{E}_{s} =184.68$. This broadening is a signature of inclusion of emission of multiple phonons, referred to as acoustic phonon-broadening\cite{besombes,Sengupta2017,jakovac,jakovac2,jakovac3,qed1}. The peak of the Lorentzian shows significant shift related to the effects of the acoustic polaron leading to renormalization of the bound state energy. With decreasing $\tilde{\epsilon}$, we also observe enhanced broadening of the peak which is a result of increased damping related to the decay of phonon coherence.}
 \label{fig:spectral1}
\end{centering}
\end{figure}

\begin{figure}[t]
\begin{centering}
\includegraphics[width=\columnwidth]{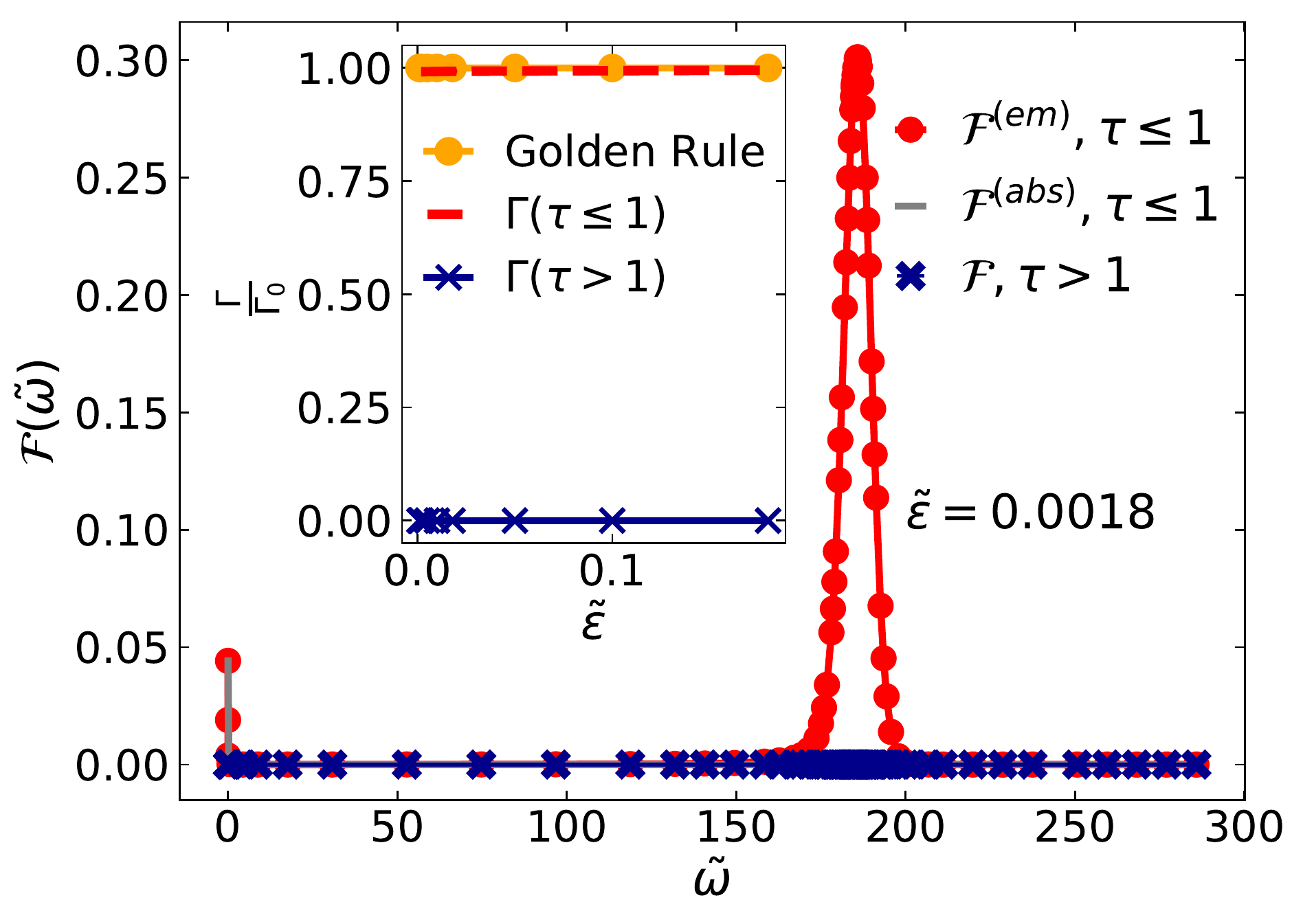}
\caption{Spectral weight distribution for the emission  $\mathcal{F}^{(em)}$ and absorption processes $\mathcal{F}^{(abs)}$ (grey line) for the two time regimes $0\leq\tau\leq1$ (red dots) and $1\leq\tau\leq\infty$ (blue crosses) for a 10 $\mu$m membrane ($\tilde\varepsilon = 0.0018$). Loss of phonon dynamics in the regime $1\leq\tau\leq\infty$ leads to a complete disappearance of the spectral weight $\mathcal{F}$ around $\tilde{\omega} =\tilde{E}_{s}$ which otherwise is present in the short time regime $0\leq\tau\leq1$ ($\mathcal{F}^{(em)}$ red dots). While a vanishingly small contribution to the adsorption rate $\Gamma(\tau>1)$ is observed in the asymptotically long time regimes (inset plot blue crosses), the short-time contribution $\Gamma(\tau\leq1)$ (red line) matches with the golden rule result (orange line). The overall many body adsorption rate $\Gamma$ which is a sum of contribution from both the time regimes matches with the golden rule result validating the Bloch-Nordsieck sum rule. }

\label{fig:spectral}
\end{centering}
\end{figure}
We calculate the respective integrals numerically and plot the variation of  $\mathcal{F}^{(em)}$ as a function of $\tilde{\omega}$ for the different values of $\tilde{\epsilon}$ in Fig.~\ref{fig:spectral1}. For all three values of $\tilde{\epsilon}$, we notice a broad peak around the non-renormalized bound state energy $\tilde{\omega} =\tilde{E_{s}} = E_{s}/\sqrt{g_{b}^{2}T} = 184.68$ (denoted by vertical black dashed line). The salient features associated with these lorentzian curves are as follows: (i) the peaks do not appear exactly at $\tilde{\omega} = \tilde{E}_{s}$ but are shifted and we relate this shift to the polaronic effects associated with the renormalization of the bound state energy. Also, the respective values of the shift due to different $\tilde{\epsilon}$ are negligible compared to each other, as expected (see Fig.~\ref{fig:renorme}, where the renormalization effects show almost no variation with $\tilde{\epsilon}$ for $0\leq\tau\leq1$). (ii) We observe damping of the curves and relate this to the effect of the decay of the bound state propagator which represents the dephasing of the phonons. We observe an enhanced broadening due to increased damping with decrease in $\tilde{\epsilon}$ which can be understood from the rapid loss of coherence of the phonons with decreasing $\tilde{\epsilon}$ (see the dependence of $\mathcal{\tilde{S}}$ on $\tilde{\epsilon}$ for $\tau<1$ in  Fig.~\ref{fig:decaye}). The broadening of the function $\mathcal{F}^{(em)}$ is a signature of the effect of inclusion of emission of multiple phonons and is also observed in other branches of quantum field theories which use resummations involving multiple quasiparticles \cite{qed,qed1, jakovac,jakovac2,jakovac3}. 

While we observe an accumulation of spectral weight around $\tilde{\omega} = \tilde{E}_{s}$, there is also a loss in the magnitude of the weight with decreasing $\tilde{\epsilon}$. Naturally, we ask if this loss in spectral weight around $\tilde{\omega} \approx \tilde{E_{s}}$ re-appears at a different phonon frequency? To answer this, we look at the total emission and absorption spectra for $\tilde{\epsilon} =0.0018$ corresponding to the entire phonon frequency scale (see Fig.~\ref{fig:spectral}). Quite interestingly, we observe in addition to the spectral weight around $\tilde{\omega} \approx \tilde{E}_{s}$ for $\mathcal{F}^{(em)}$, there appears an accumulation of weight around $\tilde{\omega} = \tilde{\epsilon}$. For higher values of $\tilde{\epsilon}$, we have checked that the spectral weights around $\tilde{\omega} =\tilde{\epsilon}$ are negligible and increases with decreasing $\tilde{\epsilon}$ and even for the lowest value of $\tilde{\epsilon} = 0.0018$ ($L =10\mu$m) this is still less in magnitude compared to the spectral weight around $\tilde{\omega}\approx \tilde{E}_{s}$. For the absorption spectra given by $\mathcal{F}^{(abs)}$, there is no appearance of spectral weight around $\tilde{\omega}\approx \tilde{E}_{s}$ but only around $\tilde{\omega} = \tilde{\epsilon}$. This suggests that the adsorption phenomenon is indeed mediated by the emission of phonons of 2 definite frequencies: a hard phonon corresponding to $\tilde{\omega} \sim \tilde{E}_{s}$ and a soft phonon $\tilde{\omega} \sim \tilde{\epsilon}$, additionally the renormalization of bound state energy and decay of the propagator is totally controlled by the emitted phonons. Let us now look at the contribution from the other regime $1<\tau\leq\infty$.

\subsection{Contribution from regime $1<\tau\leq\infty$}
\label{sec:longtimes}
We follow the same procedure as before but subject Eqs.~\ref{FSimagG},~\ref{IMSKK},~\ref{Fem} and ~\ref{Fabs} to condition $1<\tau\leq\infty$ (i.e, integrate the transformed Eq.~\ref{FSimagG} for regime $1\leq\tau\leq\infty$). Let us begin by looking at the variation of $\mathcal{F} = \mathcal{F}^{(em)} + \mathcal{F}^{(abs)}$ as a function of  $\tilde{\omega}$. As we noticed from the previous subsection, the contribution from the soft phonon emission is visible only for $\tilde{\epsilon} =0.0018$, hence we show our calculations in this regime for this lowest value. We find the contribution from $\mathcal{F}^{(abs)}$ to $\mathcal{F}$ to be negligible (similar to regime $0\leq\tau\leq1$), however we observed a surprising feature in the contribution from $\mathcal{F}^{(em)}$. In Fig.~\ref{fig:spectral}, we note a complete disappearance of the spectral weight around $\tilde{\omega} \approx \tilde{E}_{s}$. To understand this feature, we recall this regime is characterized by massive renormalization effects with complete loss of phonon coherence. In fact it is the vanishing of the decay factor $\tilde{S}$ represented by the shift due to the Franck-Condon effect that results in the complete absence of the spectral weight at $\tilde{\omega}\approx \tilde{E}_{s}$ (see the dependence of $\mathcal{\tilde{S}}$ on $\tilde{\epsilon}$ for $\tau>1$ corresponding to $\tilde{\epsilon}<0.5$ in Fig.~\ref{fig:decaye}). Quite naturally, the contribution from $\mathcal{F}^{(em)}$ and $\mathcal{F}^{(abs)}$ to $\mathcal{F}$ is negligibly small in this regime and tends to 0 without any appreciable accumulation of spectral weight at $\tilde{\omega}\sim\tilde{\epsilon}$ or $\tilde{\omega}\sim\tilde{E}_{s}$ (see Fig.~\ref{fig:spectral}).

Let us now calculate the final many-body adsorption rate for the full time regime $0\leq\tau\leq\infty$ which comprises the contribution from the regimes $0\leq\tau\leq1$ and $1<\tau\leq\infty$.

\subsection{Final Adsorption rate for the full time regime $0\leq\tau\leq\infty$}
\label{sec:fulltime}

In this section, we discuss the final adsorption rate which is a sum of the contribution from the regimes $0\leq\tau\leq1$ and $1<\tau\leq\infty$,
\begin{equation}\label{Gammafinal}
\Gamma(0\leq\tau\leq\infty) = \Gamma(\tau\leq1) +\Gamma (\tau>1)
\end{equation}
We calculate $\Gamma$ following Eq.~\ref{TRIBM} using Eqs.~\ref{FSimagG}, \ref{IMSKK}, \ref{Fem} and \ref{Fabs} for both the regimes $\tau\leq 1$ and $\tau>1$. For comparative purposes we normalize the final rate $\Gamma$ with respect to the Golden rule result $\Gamma_{0}$ given by Eq.~\ref{GR} and show the variation of $\Gamma/\Gamma_{0}$ with respect to the dimensionless IR cut-off $\tilde{\epsilon}$ in the inset of Fig.~\ref{fig:spectral}. For a discussion on the quasiparticle weight $\mathcal{Z}$ that appears in the calculation of $\Gamma$, see Appendix Sec.~\ref{sec:Z}. 

For membrane sizes $10~\mu$m $\leq L \leq 100$ nm (corresponding to IR cut-off $0.0018\leq\tilde{\epsilon}<0.2$) and maintained at T = 10 K, we plot the contribution from the regime $\tau\leq 1$ shown in the inset by $\Gamma(\tau\leq1)$ (red dashed lines) and find it to match the conventional golden rule result $\Gamma_{0}$. However, for the same membrane parameters, the contribution from $\Gamma(\tau>1)$ corresponding to regime $\tau>1$ is found to be 0 (blue crosses in the inset of Fig.~\ref{fig:spectral}). Thus the total many-body adsorption rate is seen to be dominated by the contribution from the regime $0\leq\tau\leq 1$ and can be written as
\begin{equation}\label{g}
\Gamma(0\leq\tau\leq\infty) \approx \Gamma(\tau\leq1)\approx \Gamma_{0}
\end{equation}

We understand this result by recalling the variation of the spectral weight $\mathcal{F}(\tilde{\omega})$ for all frequencies $\tilde{\omega}$ for both the emission and absorption processes. In fact, the total adsorption rate is equivalent to summing the spectral weights over all frequencies (see Eqs.~\ref{IMSKK},~\ref{spec},~\ref{Fem} and \ref{Fabs}.) Since in regime $\tau>1$, there is a complete absence of spectral weight around $\tilde{\omega} \sim \tilde{E}_{s}$ due to the complete loss of phonon coherence which leads to the emergence of the Franck-Condon shift, the contribution to $\Gamma$ from $\Gamma(\tau>1)$ is naturally 0. However, for the regime $\tau\leq1$, the renormalization effect from the emitted thermal phonons leads to a broadened density of state around $\tilde{\omega}\sim \tilde{E}_{s}$ with a small contribution from the soft phonons at $\tilde{\omega}\sim\tilde{\epsilon}$, hence when summed over all frequencies, this regime contributes to the adsorption rate. As a matter of fact, the total adsorption rate is seen to be dominated by the effects from the contributions from the regime $\tau\leq1$. 

From Eq.~\ref{g}, we see that the final many-body adsorption rate after the resummation procedure has resulted in a rate that is equal to the Golden rule result $\Gamma_{0}$ which indicates that the rate is indeed independent of the contribution from soft-phonons $\epsilon$ and atom-phonon coupling in the bound $g_{b}^{2}$ (see inset of Fig.~\ref{fig:spectral} where the red dashed lines indicate the final many-body adsorption rate). This result although surprising (since we have summed over multiple contributions from soft-phonons $\epsilon$ and included interactions in all orders of $g_{b}^{2}$ in the bound state propagator) represents the essence of the Bloch-Nordsieck resummation technique. According to the Bloch-Nordsieck sum rule (theorem), the final scattering rate after summing over all soft quanta is found to be identical to the cross-section of scattering in absence of any interaction with the radiation field \cite{Bloch:1937pw,rosenberg1,rosenberg2,ALTHERR1990360}. Hence, Eq.~\ref{g} validates the Bloch-Nordsieck sum rule for our model. Our previous result for zero temperature \cite{Sengupta2017} was also seen to satisfy the sum rule. Nevertheless, also as a check of our result for the many-body adsorption rate, we verified the sum rule that the bound state propagators must obey\cite{mahan},
\begin{equation}\label{sum}
1= -\frac{1}{\pi}\int_{\tilde{\epsilon}}^{\tilde{\omega_{D}}} \bigg\{\mathcal{I}m \tilde{G}_{bb}(\tilde{E_{s}}-\tilde{\omega}) \bigg\} \mathrm{d}\tilde{\omega}\\
\end{equation}
For the full time regime given by $0\leq\tau\leq\infty $, the numerically calculated propagator corresponding to the transformed version of Eq.~\ref{FSimagG} was found to obey the above defined sum rule (Eq.~\ref{sum}).

\section{Conclusion}
\label{sec:conclusion}
In summary, we have modelled the phenomenon of adsorption on suspended graphene membranes based on an exactly solvable Hamiltonian in many-body physics, the independent boson model (IBM). The success of IBM traces back to the exact treatment of phonon dynamics which allows us to investigate in detail the role played by the multiphonon emission and absorption in assisting the adsorption of an atom to a membrane.

While our simple model of adsorption based on the IBM helps us to understand the renormalization of the physisorption well, relating it to the formation of a $\textit{phonon-dressed}$ atom which we refer to as the acoustic polaron, it also sheds some light on the decay of the atom in terms of the time-evolution of the phonon bath correlator. This in particular turns out to be important. Our study finds the decay of the bound atom propagator is strongly dependent on the time-evolution of the phonon bath correlations with surprisingly contrasting results corresponding to different regimes of a dimensionless characteristic time scale $\tau$, defined as $\tau=(\sqrt{g_{b}^{2}T})t =1$ with $t$, $g_{b}^{2}$ and $T$ as the time, atom-phonon interactions and membrane temperature, respectively. With an aim to understand this dependence, we constructed a new dimensionless IR scale $\tilde{\epsilon} = \epsilon/\sqrt{g_{b}^{2}T}$ which comprises the effects of temperature, soft phonon ($\epsilon$) contribution  (equivalent to the effect of finite-sized membranes) and the atom-phonon interactions. Corresponding to various values of $\tilde{\epsilon}$, we indeed find phonons to exhibit different dynamics in the regimes $\tau<1$ and $\tau>1$. While for micromembranes, phonons exhibit a short-time dynamics with rapid loss of coherence in the regime $\tau<1$, for membranes with sizes $<$ 100 nm, phonons exhibit long-time dynamics with negligible loss of coherence. We also understand the long-time result ($\tau>1$) as an indication of the emergence of the shift in the phonon bath correlator due to the time-independent Franck-Condon factor. Thus our study reveals well-defined time scales for infrared phonon dynamics which in principle can be controlled by the atom-phonon interaction strength and membrane-temperature.

For micromembranes we computed the adsorption rates based on a resummation technique that uses the exact solution of the IBM as the dressed propagator for the atom self-energy. The contribution from the different regimes of $\tau$ towards the many-body adsorption rate $\Gamma$ were found to be completely contrasting. For $0\leq\tau\leq1$ i.e, $0\leq t\leq(g_{b}^{2}T)^{-1/2}$ and as a matter of fact, for the full time regime $0\leq t\leq\infty$, we derived adsorption rates that are finite, approximately equal to the Golden rule result and shows negligible dependence on the IR cut-off (or the soft-phonon emission). This is in agreement with the Bloch-Nordsieck sum rule which predicts that the resummed scattering rate including the summation of contribution of emission of infinitely many soft quanta is identical to the scattering rate calculated in the absence of any interaction with the radiation field\cite{Bloch:1937pw,rosenberg1,ALTHERR1990360}. While the main contribution to the many-body adsorption rate came from regime $0\leq t\leq(g_{b}^{2}T)^{-1/2}$, the contribution from $1<\tau\leq\infty$ or $(g_{b}^{2}T)^{-1/2}\leq t\leq\infty$ was found to be zero due to the complete loss of phonon coherence leading to the Franck-Condon shift.

We thus conclude that for suspended graphene micromembranes maintained at 10 K or sufficiently low temperature (temperature much less than the physisorption well energy but greater than the atom-phonon coupling and IR scale), the many-body adsorption rate will be finite, equivalent to the golden rule result and independent of the size of the membrane which implies that the effect of low-energy phonons to the adsorption rate is negligible (see Eq.~\ref{g} and inset of Fig.~\ref{fig:spectral}). Our results are thus in agreement with Ref.~[\onlinecite{LJreply,LJ2011}] and disagree with Ref.~[\onlinecite{clougherty2017T}]. In Ref.~[\onlinecite{clougherty2017T}], it seems to us that the many-body adsorption rate has been derived under an approximation that might have neglected the contribution from the effects of emission of thermal phonons.  As our study indicates, it is in fact crucial to incorporate the effects of the thermal phonon emission which if neglected would give an adsorption rate that is zero (see Fig.~\ref{fig:spectral}) but would violate the sum rule (defined by Eq.~\ref{sum}). Thus it would be interesting to see if the result for the adsorption rate in Ref.~[\onlinecite{clougherty2017T}] would change with the inclusion of the effects of the thermal phonons. We must also mention a difference in the ``trapping mechanism" between our work and Ref.~[\onlinecite{LJ2011}], where the authors have considered a ``diffraction mediated selective adsorption resonance" which allows the hydrogen atom more time to exchange low-frequency phonons with the surface before getting stuck. This mechanism gives rise to an additional enhancement in sticking probabilities for incident energies 7 meV - 15 meV. Since our current work does not include this mechanism, we do not observe this additional enhancement. We leave this investigation for future work. We envision our methodology and results to apply to other 2D materials with sufficiently weak atom-phonon coupling, thus opening up the possibility of application of these materials as nano-mechanical devices used as mass sensors\cite{Garcia-Sanchez2008,LJ2011}.

While we started this paper with a question, let us also end with an analogy and a few pertinent questions related to the IR divergence in other field-theories. As mentioned before in the Introduction (Sec.~\ref{sec:intro}), within the theories of QED and perturbative gravity, the massless nature of photons and gravitons lead to straightforward divergences in the perturbation series for scattering rates due to the emission of low-energy virtual bosons \cite{Bloch:1937pw,weinberg}. The solution to this IR catastrophe is then realized within resummation schemes using formalisms that were first developed in the field of electrodynamics by Bloch-Nordsieck \cite{Bloch:1937pw} and in gravity by Weinberg\cite{weinberg}. The results of these resummations in QED and perturbative gravity led to the emergence of $\textit{Soft theorems}$ which relate the matrix elements of a Feynman diagram with an external soft quanta insertion to that of the same diagram without an external soft quanta \cite{weinberg_1995}. In recent years, there has been a renewed interest to understand the IR structure of these theories with a motivation to decipher the connections between the vividly disparate fields of soft theorems and the information theoretic properties of the soft radiations\cite{carney,carneyprd,Gómez2018,Strominger}. It is in fact believed that for every massless quanta there exists a connection between the soft theorems (IR catastrophe), memory effects and asymptotic symmetries\cite{Strominger}. In our simple model of a low-energy atom interacting with a suspended graphene membrane, we encounter similar IR divergent adsorption rates due to the emission of infinitely many soft phonons. We employ the Bloch-Nordsieck scheme of resummation formalism employed in QED \& perturbative gravity\cite{Bloch:1937pw,weinberg} which amounts to including the emissions of infinitely many soft phonons and derive a non-perturbative result for the many-body adsorption rate which is finite and independent of the soft phonon contribution thereby validating the Bloch-Nordsieck theorem\cite{Bloch:1937pw}. Thus there is a subtle similarity between the theories of QED, perturbative gravity and our model in terms of the IR problem, the technique to address the IR problem and finally the nature of the solution to the problem. Thus motivated by these similarities we ask, would the soft phonon radiation encode information? 

\section*{acknowledgments}
I thank Professors Dennis P. Clougherty and Valeri N. Kotov for stimulating discussions. I am grateful to Professor Ion Garate for his valuable advice, insightful comments and suggestions. I thank Samuel Boutin and Benjamin Himberg for their questions and comments. This work was funded by the Canada First Research Excellence Fund.

\appendix
\section{Real Atom self-energy within IBM and quasiparticle weight Z}
\label{sec:Z}

\begin{figure}
    \centering
    \includegraphics[width=0.95\columnwidth]{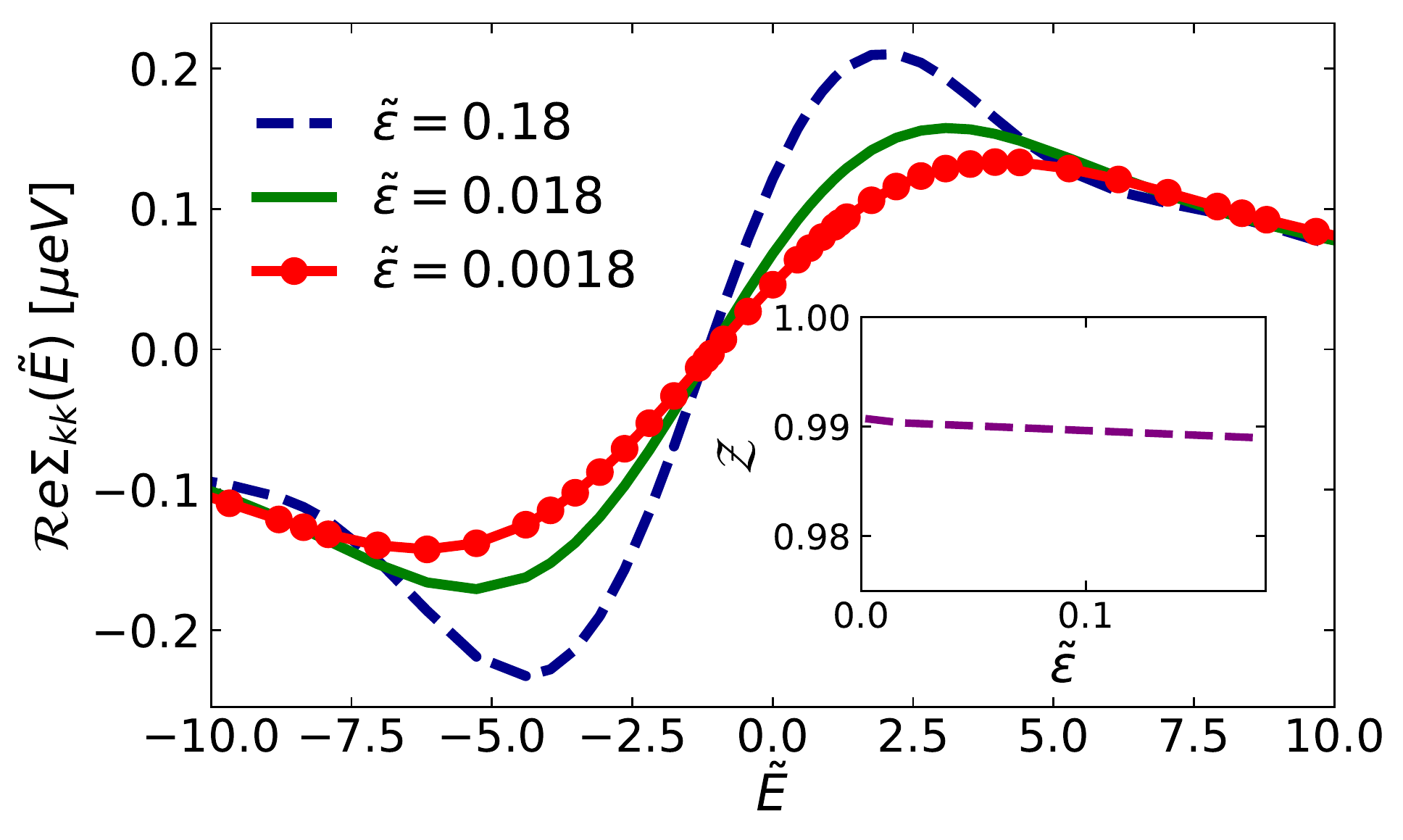}
    \caption{The real part of the atom self-energy is plotted as a function of various values of the dimensionless IR cut-off $\tilde{\epsilon}$. Inset shows the variation of the quasiparticle weight $\mathcal{Z}$. We note $\mathcal {Z} \leq 1$ for the chosen values of $\tilde{\epsilon}<0.2$. }
    \label{fig:Z}
\end{figure}

Using our method of resummation as given in Sec.~\ref{sec:prelim}, we write the real part of the atom self-energy as:
\begin{equation}\label{realSkk}
\begin{split}
\mathcal{R}e\Sigma_{kk}^{(IBM)} &= g_{k}^{2}\bigg[\int_{\tilde{\epsilon}}^{\tilde{\omega_{D}}}\mathrm{d}\tilde{\omega}\bigg\{\frac{1}{\exp(\tilde{\omega}\sqrt(g_{b}^{2}/T)-1}+1\bigg\}\\
&\quad\mathcal{R}e\tilde{G}_{bb}(\tilde{E}_{s}-\tilde{\omega})\bigg]\\
&\quad + g_{k}^{2}\bigg[\int_{\tilde{\epsilon}}^{\tilde{\omega_{D}}}\mathrm{d}\tilde{\omega}\bigg\{\frac{1}{\exp(\tilde{\omega}\sqrt(g_{b}^{2}/T)-1}\bigg\}\\
&\quad\mathcal{R}e\tilde{G}_{bb}(\tilde{E}_{s}+\tilde{\omega})\bigg]\\
\end{split}
\end{equation}
where, we have used the transformations defined by Eqs.~\ref{trans},~\ref{limits} and~\ref{time} given in Sec.~\ref{sec:phononcorr}. Our aim in this section, would be to calculate the real part of $\Sigma_{kk}^{(IBM)}$ and derive the respective values of the quasiparticle weight $\mathcal{Z}$ (given by Eq.~\ref{RF}) for membrane sizes 100 nm to 10 $\mu$m maintained at 10 K.

We accomplish this by numerically integrating Eq.~\ref{realSkk} using the expression of the real part of the bound state propagator $\mathcal{R}eG_{bb}$ given by Eq.~\ref{FSrealG} (transformed accordingly) for $\tilde{\epsilon} =0.0018, 0.018$ and 0.18. In Fig.~\ref{fig:Z}, we plot the variation of the $\mathcal{R}e\Sigma_{kk}(\tilde{E})$ vs $\tilde{E}$. We have numerically checked that the real part of $\Sigma_{kk}^{(IBM)}(\tilde{E})$ vanishes approximately at the energy corresponding to the peak in the $\mathcal{I}m\Sigma_{kk} (\tilde{E})$. The function is well-behaved for the chosen values of $\tilde{\epsilon}$ and the corresponding value of the quasiparticle weight was evaluated numerically and found as $\mathcal{Z}\leq 1$ without any significant dependence on $\tilde{\epsilon}$ (see inset of Fig.~\ref{fig:Z}).

\bibliographystyle{apsrev4-1}
\bibliography{rudraads}

\end{document}